\documentclass{aa}
\usepackage{graphicx}
\usepackage{txfonts}
\newcommand{\iacs}{ACS/WFC 3$\times$3 mosaic}
\newcommand{\wfi}{WFI@2.2$m$}

\newcommand{\re}{$r^*$}
\newcommand{\rc}{$r_{c}$}
\newcommand{\rh}{$r_{h}$}
\newcommand{\nbr}{$N_{\rm{bMS}}/N_{\rm{rMS}}$}
\newcommand{\om}{\hbox{{$\omega$}\ Cen~}}
\newcommand{\oma}{\hbox{{$\omega$}\ Cen}}

\newcommand{\masyr}{$\rm{mas}\,\rm{yr}^{-1}$}

\begin{document}
\title{Radial distribution of the multiple stellar populations\\
in $\omega$\ Centauri
\thanks{Based on observations with the NASA/ESA {\it Hubble Space
Telescope}, obtained at the Space Telescope Science Institute, which
is operated by AURA, Inc., under NASA contract NAS 5-26555, and on
observations made with ESO telescopes at La Silla and Paranal
Observatories.}}

\author{Bellini, A.\inst{1,}\inst{2}
\and
Piotto, G.\inst{1}
\and
Bedin, L.~R.\inst{2}
\and
King, I.~R.\inst{3}
\and
Anderson, J.\inst{2}
\and
Milone, A.~P.\inst{1}
\and
Momany, Y.\inst{4}
}
\offprints{Bellini, A.}
\institute{ Dipartimento di Astronomia, Universit\`a di Padova, Vicolo
dell'Osservatorio 3, I-35122 Padua, Italy\\
\email{[andrea.bellini;giampaolo.piotto;antonino.milone]@unipd.it}
\and Space Telescope Science Institute, 3700 San Martin Drive,
Baltimore, MD 21218, USA\\ \email{[bellini;bedin;jayander]@stsci.edu}
\and Department of Astronomy, University of Washington, Seattle, WA
98195-1580, USA\\ \email{king@astro.washington.edu} \and INAF:
Osservatorio Astronomico di Padova, vicolo dell'Osservatorio 5, 35122
Padova, Italy\\ \email{yazan.almomany@oapd.inaf.it} }

\date{Received 24 June 2009 / Accepted 23 September 2009}

%#####################################################################

\abstract {} {We present a detailed study of the radial distribution
of the multiple populations identified in the Galactic globular
cluster $\omega$\ Cen.}  {We used both space-based images (ACS/WFC and
WFPC2) and ground-based images (FORS1@VLT and WFI@2.2$m$ ESO
telescopes) to map the cluster from the inner core to the outskirts
($\sim$20 arcmin).  These data sets have been used to extract
high-accuracy photometry for the construction of color-magnitude
diagrams and astrometric positions of $\sim900\,000$ stars.}  {We find
that in the inner $\sim$2 core radii the blue main sequence (bMS)
stars slightly dominate the red main sequence (rMS) in number.  At
greater distances from the cluster center, the relative numbers of bMS
stars with respect to rMS drop steeply, out to $\sim$8 arcmin, and
then remain constant out to the limit of our observations.  We also
find that the dispersion of the Gaussian that best fits the color
distribution within the bMS is significantly greater than the
dispersion of the Gaussian that best fits the color distribution
within the rMS.  In addition, the relative number of
intermediate-metallicity red-giant-branch stars which includes the
progeny of the bMS) with respect to the metal-poor component (the
progeny of the rMS) follows a trend similar to that of the
main-sequence star-count ratio \nbr.  The most metal-rich component of
the red-giant branch follows the same distribution as the
intermediate-metallicity component.}  {We briefly discuss the possible
implications of the observed radial distribution of the different
stellar components in $\omega$~Cen.}


\maketitle

%#####################################################################

\section{Introduction}
\label{sec:introduction}

The globular cluster (GC) $\omega$\ Centauri is the most-studied
stellar system of our Galaxy, but nevertheless one of the most
puzzling.  Its stars cover a wide range in metallicity (Cannon \&
Stobie \cite{cannon73}; Norris \& Bessell \cite{norris75},
\cite{norris77}; Freeman \& Rodgers \cite{freeman75}; Bessell \&
Norris \cite{bessell76}; Butler et al.\ \cite{butler78}; Norris \& Da
Costa \cite{norris95}; Suntzeff \& Kraft \cite{sunt96}; Norris et al.\
\cite{norris96}), with a primary component at [Fe/H] $\sim-1.7$ to
$-1.8$, and a long tail extending up to [Fe/H] $\sim-0.6$, containing
three or four secondary peaks (see Johnson et al.\ \cite{johnson09}
for a recent update).  It has been shown, both with ground-based
photometry (Lee et al.\ \cite{lee99}; Pancino et al.\
\cite{pancino00}; Rey et al.\ \cite{rey04}; Sollima et al.\
\cite{sollima05a}; Villanova et al.\ \cite{villanova07}) and
\textit{Hubble Space Telescope} (\textit{HST}) photometry (Anderson
\cite{anderson97}; Bedin et al.\ \cite{bedin04}; Ferraro et al.\
\cite{ferraro04}), that \om\ hosts different stellar populations, most
of them clearly visible in most of their evolutionary phases.

These populations have been linked to the aforementioned metallicity
peaks, in photometric studies of the red-giant branch (RGB) (Pancino
et al.\ \cite{pancino00}; Hilker \& Richtler \cite{hilker00}; Sollima
et al.\ \cite{sollima05a}), the subgiant branch (SGB) (Hilker et al.\
\cite{hilker04}; Sollima et al.\ \cite{sollima05b}; Stanford et al.\
\cite{stanford06}; Villanova et al.\ \cite{villanova07}), and the main
sequence (MS) (Piotto et al.\ \cite{piotto05}).  The most puzzling
feature in \om was discovered by Piotto et al.\ (\cite{piotto05}), who
showed that, contrary to any expectation from stellar-structure
theory, the bluer of the two principal main sequences (bMS) is more
metal-rich than the redder one (rMS).  The only possible way of
reconciling the spectroscopic observations with the photometric ones
is to assume a high overabundance of He for the bluer MS (Bedin et 
al.\ \cite{bedin04}; Norris \cite{norris04}; Piotto et al.\
\cite{piotto05}).  How such a high He content could have been formed
is still a subject of debate (see Renzini \cite{renzini08} for a
review).

One of the scenarios proposed to account for all the observed features
of \om is a tidal stripping of an object that was originally much more
massive (Zinnecker et al.\ \cite{zinnecker88}; Freeman
\cite{freeman93}; Dinescu et al.\ \cite{dinescu99}; Ideta \& Makino
\cite{ideta04}; Tsuchiya et al. (\cite{tsuchiya04}); Bekki \& Norris
\cite{bekki06}; Villanova et al.\ \cite{villanova07}).  In
this scenario, the cluster was born as a dwarf elliptical galaxy,
which was subsequently tidally disrupted by the Milky Way.  Since all
the populations of such a galaxy pass through the center, the nucleus
would have been left with a mixture of all of them.

It has also been suggested (Searle \cite{searle77}; Makino et al.\
\cite{makino91}; Ferraro et al.\ \cite{ferraro02}) that \om\ could
have been formed by mergers of smaller stellar systems.  In apparent
support of this scenario, Ferraro et al.\ (\cite{ferraro02}) claimed
that the most metal-rich RGB component of \om\ (RGB-a, following the
nomenclature of Pancino et al.\ \cite{pancino00}) has a significantly
different mean proper motion from that of the other RGB stars, and
they concluded that RGB-a stars must have had an independent origin.
However, Platais et al.\ (\cite{platais03}) showed that the
proper-motion displacement seen could instead be an uncalibratable
artifact of the plate solution.  More recently Bellini et al.\
(\cite{bellini09}), with a new CCD-based proper-motion analysis, were
able to demonstrate that all \om\ RGB stars share the same mean motion
to within a few km/sec.  Anderson \& van der Marel
(\cite{anderson09b}) also find that the lower-turnoff population (the
analog of the RGB-a) shows the same bulk motion as the rest of the
cluster.  Thus there is no longer a reason to think this population is
kinematically distinct and an indication of a recent merger.  Another
indication that the cluster likely did not form by mergers can be
found in the observation in Pancino et al.\ (\cite{pancino07}) that
all three RGB components share the cluster rotation, which would not
be the case if different populations had different dynamical origins,
or at least would require an unlikely degree of fine tuning.

While $\omega$ Cen was long thought to be the only cluster to exhibit
a spread in abundances, we now know that it is not alone.  M54 also
clearly exhibits multiple RGBs (Sarajedini \& Layden
(\cite{ata95}); Siegel et al.\ \cite{siegel07}), SGBs (Piotto
\cite{piotto09}), and has hints of multiple MSs.  The complexity of
M54 makes good sense, because it coincides with the nucleus of the
tidally disrupting Sagittarius dwarf-spheroidal galaxy.  M54 might be
the actual nucleus or, more likely, it may represent a cluster that
migrated to the nucleus as a result of dynamical friction (Bellazzini
et al.\ \cite{bellazzini08}).  $\omega$ Cen and M54 are the two most
massive GCs in our Galaxy, and it is quite possible that they are the
result of similar---and peculiar---evolutionary paths (Piotto
\cite{piotto09}).  In any case, even \om and M54 are not the only
clusters to exhibit non-singular populations.  Exciting new
discoveries, made in the last few years, clearly show that the GC
multi-population zoo is quite populated, inhomogeneous, and complex.

Piotto et al.\ (\cite{piotto07}) published a color-magnitude diagram
(CMD) of the globular cluster NGC 2808, in which they identified a
well-defined triple MS (D'Antona et al.\ [\cite{dantona05}] had
already suspected an anomalous broadening of the MS and had associated
it with the three populations proposed by D'Antona \& Caloi
[\cite{dantona04}] to explain the complex horizontal branch (HB) of
this cluster).  Another globular cluster, NGC 1851, must have at least
two distinct stellar populations. In this case, the observational
evidence comes from the split of the SGB (Milone et al.\
\cite{milone08}).  There are other GCs which undoubtedly show a split
in the SGB, like NGC 6388 (Moretti et al.\ \cite{moretti09}), M22
(Piotto \cite{piotto09}; Marino et al.\ (\cite{marino09}), 47 Tuc
(Anderson et al.\ \cite{anderson09a}), which also shows a MS
broadening, or in the RGB, like M4 (Marino et al.\ \cite{marino08}).
Recent investigations (Rich et al. \cite{rich04}; Faria et
al. \cite{faria07}) suggest that also other galaxies might host GCs
with more than one population of stars.

Multiple-population GCs offer observational evidence that challenges
the traditional view.  For half a century, a GC has been considered to
be an assembly of stars that (quoting Renzini \& Fusi Pecci
\cite{renzini88}): ``\textit{represent the purest and simplest stellar
populations we can find in nature, as opposed to \emph{field}
populations, which result from an admixture of ages and
compositions}''.  If we allow for the fact that all the GCs for which
Na and O abundances have been measured show a well defined Na/O
anti-correlation (Carretta et al.\ \cite{carretta06},
\cite{carretta08}), suggesting an extended star-formation process, and
that 11 of the 16 intermediate-age Large Magellanic Cloud GCs have
been found to host multiple populations (Milone et al.\
\cite{milone09}), multi-populations in GCs could be more the rule than
the exception.  {\it De facto}, a new era in globular-cluster research
has started, and understanding how a multiple stellar system like \om
was born and has evolved is no longer the curious study of an anomaly,
but rather may be a key to understanding basic star-formation
processes.

One way to understand how the multiple populations may have originated
is to study the spatial distributions of the different populations,
which might retain information about where they formed.  In
particular, theoreticians have been finding that if the second
generation of stars is formed from an interstellar medium polluted
and shocked by the winds of the first generation, then we would expect
that the second generation would be more concentrated towards the
center of the cluster than the first one (see D'Ercole et al.\
\cite{d'ercole08}; Bekki \& Mackey \cite{bekki09}; Decressin et al.\
\cite{decressin08}).  In the last of these references it is shown that
in such a scenario the two generations of stars would interact
dynamically and would homogenize their radial distributions over time.
As such, spatial gradients represent a fading fossil record of the
cluster's dynamical history.

Since $\omega$ Cen has such a long relaxation time (1.1 Gyr in the
core and 10 Gyr at the half-mass radius, Harris \cite{harris96}), it
is one of the few clusters where we might hope to infer the
star-formation history by studying the internal kinematics and spatial
distributions of the constituent populations.  These measurements will
provide precious hints and constraints to allow theoreticians to
develop more reliable GC dynamical models.

In a recent paper, Sollima et al.\ (\cite{sollima07}) showed that the
star-count ratio \nbr\ is flat beyond $\sim12\arcmin$, but that inward
to $\sim8\arcmin$ it increases to twice the envelope value.  Thus the
bMS stars (i.e., the supposed ``He-enriched'' population) are more
concentrated towards the center than the rMS, which is presumed to be
the first generation.  Unfortunately, Sollima et al.\
(\cite{sollima07}) could not provide information about the trend of
\nbr\ within $\sim8\arcmin$, which corresponds roughly to 2 half-mass
radii (\rh).

On the other hand, the radial distribution of RGB subpopulations has
been analyzed by many authors (Norris et al.\ \cite{norris97}; Hilker
\& Richtler \cite{hilker00}; Pancino et al.\ \cite{pancino00},
\cite{pancino03}; Rey et al.\ \cite{rey04}; Sollima et
al. \cite{sollima05a}; Castellani et al.\ \cite{castellani07}; Johnson
et al.\ \cite{johnson09}).  All these works agree that the
intermediate-metallicity population (RGB-MInt) is more centrally
concentrated than the more metal-poor one (RGB-MP).  However, there is
a disagreement about the most metal-rich population (RGB-a):\ Pancino
et al.\ (\cite{pancino00}), Norris et al.\ (\cite{norris97}), and
Johnson et al.\ (\cite{johnson09}) found that the most metal-rich
stars (RGB-a) are as concentrated as the intermediate-metallicity
ones, and consequently more concentrated than the most metal-poor
stars, whereas Hilker \& Richtler (\cite{hilker00}) and Castellani et
al.\ (\cite{castellani07}) considered the RGB-a component to be the
least-concentrated population.  (Since our work in progress was
already favoring the former view over the latter, we were concerned to
reach the definitive truth of this matter).

In the present paper, we trace the radial distribution of the stars of
\om, both on the MS and in the RGB region.  Our radial density
analysis covers both the center and the outskirts of the cluster,
taking advantage of the combination of four instruments on three
different telescopes, and of our proper-motion measurements on
ground-based multi-epoch wide-field images (Bellini et al.\
\cite{bellini09}).  In Section 2 we describe in detail the photometric
data and the reduction procedures.  Section 3 presents our analysis of
the radial distribution of the stars on the two MSs.  In Section 4 we
perform an analogous study for the RGB stars.  A brief discussion
follows in Section 5.

%#####################################################################

\section{Observations and data reductions}
\label{section2}

To trace the radial distribution of the different stellar populations
in \oma, we analyzed several data sets, from four different cameras.
To probe the dense inner regions of the cluster we took advantage of
the space-based high resolving power of {\it HST}, using both the Wide
Field Channel (WFC) of the Advanced Camera for Surveys (ACS), and the
Wide Field and Planetary Camera 2 (WFPC2).  For the relatively sparse
outskirts of the cluster, we instead made use of deep archival
ground-based observations collected with the FORS1 camera of the ESO
Very Large Telescope (VLT).  In addition, to link all the different
data sets into a common astrometric and photometric reference system,
we used the Wide Field Imager (WFI) at the focus of the ESO 2.2$m$
telescope (hereafter WFI@2.2$m$).  This shallower data set was also
used to study the red-giant branch in the outskirts of the cluster.

\begin{figure}[t!]
\centering
\includegraphics[width=9.0cm,height=9.0cm]{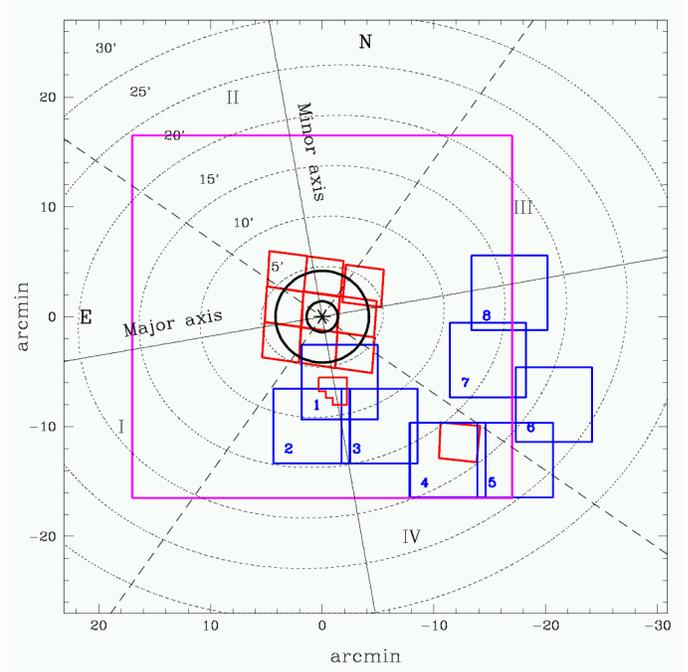}
\caption{The footprints of the \om data sets used in this work.  North
          is up, east to the left.  The ``$\ast$'' marks the cluster
          center.  The $3\times3$ ACS/WFC mosaic (in red) is that of
          GO-9442, while the 8 FORS1 fields are marked in blue.  The
          largest field (in magenta) comes from the WFI proper-motion
          catalog (Bellini et al.\ 2009).  This wide-field catalog has
          been used to register the FORS1 and the ACS/WFC inner-mosaic
          data into a common astrometric and photometric reference
          system (see text).  The smaller WFPC2 field at
          $\sim$7$^{\prime}$ south and the outer ACS/WFC field at
          $\sim$17$^{\prime}$ from the cluster center are also shown
          (in red).  Concentric ellipses, centered on the center of
          \om and aligned with the major and minor axes, show the
          radial bins that we created.  Ellipses are split into
          quadrants (dashed lines), labeled with Roman numerals.  Each
          quadrant is bisected by the major or minor axis.
	  Thick black circles mark the core radius
          ($r_c = 1\farcm4$) and the half-mass radius
          ($r_h = 4\farcm18$) (from Harris \cite{harris96}).}
\label{fig_fow}
\end{figure}

Figure \ref{fig_fow} shows the footprints of the data sets, centered
on the recently determined accurate center of \oma:\ ${\rm
RA}=13$:26:47.24, ${\rm Dec}=-47$:28:46.45 (J2000.0, Anderson \& van
der Marel \cite{anderson09b}).  The red footprints are those of {\it
HST} observations.  The larger ones are the ACS/WFC data sets, a
$3\times3$ mosaic centered on the cluster center and a single field
$\sim$17$\arcmin$ SW of the center.  The smaller red field,
$\sim$7$\arcmin$ S of the center, was observed with WFPC2.  Blue
rectangles show the partially overlapping FORS1@VLT fields, extending
from $\sim$6$\arcmin$ to $\sim$25$\arcmin$.  The large field in
magenta is the $\sim$33$\arcmin\times33\arcmin$ field-of-view of our
WFI@2.2$m$ proper-motion catalog (Bellini et al.\ \cite{bellini09}).
The figure also shows the major and minor axes (solid lines), taken
from van de Ven et al.\ (\cite{vdv06}).  We divided the field into
four quadrants, centered on the major and minor axes.  The quadrants
are labeled with Roman numerals and separated by dashed lines.  We
will use them to derive internal estimates of the errors of the
star-count distribution.  Concentric ellipses, aligned with the
major/minor axes, have ellipticity of 0.17, coincident with the
average ellipticity of \om Geyer et al. \cite{geyer83}).  These
ellipses will be used to define radial annuli, in Section 2.8.  Thick
black circles mark the core radius ($r_c = 1\farcm4$) and the
half-mass radius ($r_h = 4\farcm18$) (from Harris \cite{harris96}).
If we assuming a cluster distance of 4.7 kpc (van de Ven et al.\
\cite{vdv06}; van der Marel \& Anderson \cite{vdm09}), the two radii
correspond to 1.9 pc and 5.7 pc, respectively.

The details of the data sets are summarized in Table~\ref{tab1}.  In
the following subsections we give brief descriptions of the reduction
procedures, which have been presented in more detail in various other
papers.  The FORS1 data, however, were taken by Sollima et al.\
(\cite{sollima07}), for a purpose similar to ours; we will give a full
description of our reduction in subsection \ref{subsection2.FORS}.

\begin{table}[t!]
\centering
\caption{Data sets used in this work.}
\begin{tabular}{clr}
\hline
\hline 
&&\\ 
Data set                &Filter           & $\#$ images $\times$ Exp.\ time (s)\\
&&\\ 
\hline 
&&\\ 
$3\times3$ ACS/WFC      & F435W           & $27\times 340$, $9\times12$        \\ 
inner mosaic&&\\
                        & F625W           & $27\times 340$, $9\times8$        \\ 
&&\\ 
ACS/WFC                 & F606W           & $2\times1285$, $2\times1300$,      \\ 
$\sim$17$^\prime$       &                 & $2\times1331$, $2\times1375$      \\
                        & F814W           & $4\times1331$, $2\times1340$, $2\times1375$\\ 
&&\\ 
WFPC2@\textit{HST}      & F606W           &$2\times 300$, $1\times600$        \\ 
$\sim$7$^\prime$        & F814W           & $2\times400$, $1\times 1000$       \\ 
&&\\ 
FORS1@VLT               & $B$             & $20\times 1100$                   \\
                        & $R$             &$20\times 395$                     \\ 
&&\\ 
WFI@2.2$m$              & $B$             & $1\times10$, $1\times15$, $11\times30$, $1\times40$,\\
                        &                 & $1\times60$, $1\times120$,$2\times240$, $2\times300$\\
                        & $R_C$           & $1\times5$, $1\times10$, $1\times15$, $1\times30$, $5\times60$\\
                        & $V$             & $6\times5$, $9\times10$, $1\times15$, $3\times20$,\\
                        &                 & $2\times30$, $10\times40$, $4\times45$, $3\times60$,\\
                        &                 & $10\times90$, $7\times120$, $1\times150$, $3\times240$\\
&&\\ 
\hline
\end{tabular}
\label{tab1}
\end{table}

%#####################################################################

\subsection{\textit{HST}:\ ACS/WFC inner 3$\times$3\ mosaic}
\label{subsection2.ACS}

This data set (inner nine red rectangles in Fig.\ \ref{fig_fow},
GO-9442, PI A.\ Cool) consists of a mosaic of $3\times3$ fields
obtained with the ACS/WFC through the F435W and F625W filters.  This
camera has a pixel size of $\sim$50 mas and a field of view of
$3\farcm3\times3\farcm3$.  Each of these nine fields has one short and
three long exposures in both F435W and F625W.  The mosaic covers the
inner $\sim$10$\arcmin \times$10$\arcmin$, the most crowded region of
\oma.  These images, which were used by Ferraro et al.\
(\cite{ferraro04}) and by Freyhammer et al.\ (\cite{frey05}), and
which we used in both Bedin et al.\ ({\cite{bedin04}), and Villanova
et al.\ (\cite{villanova07}), were reduced using {\sf
img2xym\_WFC.09x10}, which is a publicly available {\sf FORTRAN}
program, described in Anderson \& King (\cite{anderson06b}).  The
program finds and measures each star in each exposure by fitting a
spatially-variable effective point-spread function.  The independent
measurements of the stars were collated into a master star list that
covers the entire $3\times3$ mosaic field.  For each star we
constructed an average magnitude in each band, and computed the rms
deviation of the multiple measurements about this average.
Instrumental magnitudes were transformed into the ACS Vega-mag flight
system following the procedure given in Bedin et al.\
(\cite{bedin05}), using the zero points of Sirianni et al.\
(\cite{sirianni05}).  Since the zero points are valid only for fluxes
in the {\tt \_drz} exposures, we computed calibrated photometry for a
few isolated stars in the {\tt \_drz} exposures and used this to set
the zero points for the photometry that was based on the individual
{\tt \_flt} images.  Saturated stars in short exposures were treated
as described in Section 8.1 in Anderson et al.\
(\cite{anderson08}). Collecting photoelectrons along the bleeding
columns allowed us to measure magnitudes of saturated stars up to 3.5
magnitudes above saturation (i.e., up to $m_{\rm F435W}$$\simeq$12
mag), with errors of only a few percent (Gilliland
\cite{gilliland04}).  We used the final catalog, which contains more
than $760\,000$ stars, to trace the radial distribution of RGB and MS
stars in this most crowded region of the cluster.

%#####################################################################

\subsection{\textit{HST}: ACS/WFC outer field}
\label{subsection2.OUT}

The outer ACS field ($\sim$17$\arcmin$ SW of the cluster center, see
Fig.~\ref{fig_fow}) comes from proposals GO-9444 and GO-10101 (both
with PI I.\ R.\ King), using the F606W and F814W filters.  The
photometry from the first-epoch observations was published in Bedin et
al.\ (\cite{bedin04}).  The photometry presented in the present paper
comes from the full two-epoch data set for this field; the two epochs
also allow us to derive proper motions and perform a critical
cluster/field separation.  A detailed description of the data
reduction, the proper-motion measurement, and the resulting CMDs
will be presented in a forthcoming paper.  The reduction and
calibration of these data sets use procedures similar to those used
for the central mosaic, and provided photometry for $\sim$3500 stars.

%#####################################################################

\subsection{\textit{HST}: WFPC2 field}
\label{subsection2.PC}

We also make use of one WFPC2 field, $\sim$7$\arcmin$ south of the
cluster center (see Fig.~\ref{fig_fow}).  This data set consists of
$2\times300+600$s exposures in F606W, and $2\times400+1000$s in F814W
(GO-5370, PI R.\ Griffiths), and contains 9214 stars.  These images
have been reduced with the algorithms described in Anderson \& King
(\cite{anderson00}).  The field was calibrated to the photometric
Vega-mag flight system of WFPC2 according to the prescriptions of
Holtzman et al.\ (\cite{holtzman95}).  This WFPC2 field is
particularly important in tracing the distribution of stars in the MS
of \oma, because it is at a radial distance from the center of the
cluster where there are no suitable ACS/WFC observations and where
ground-based observations are almost useless because of crowding.

\begin{figure}[t!]
\centering
\includegraphics[width=9.0cm,height=11.0cm]{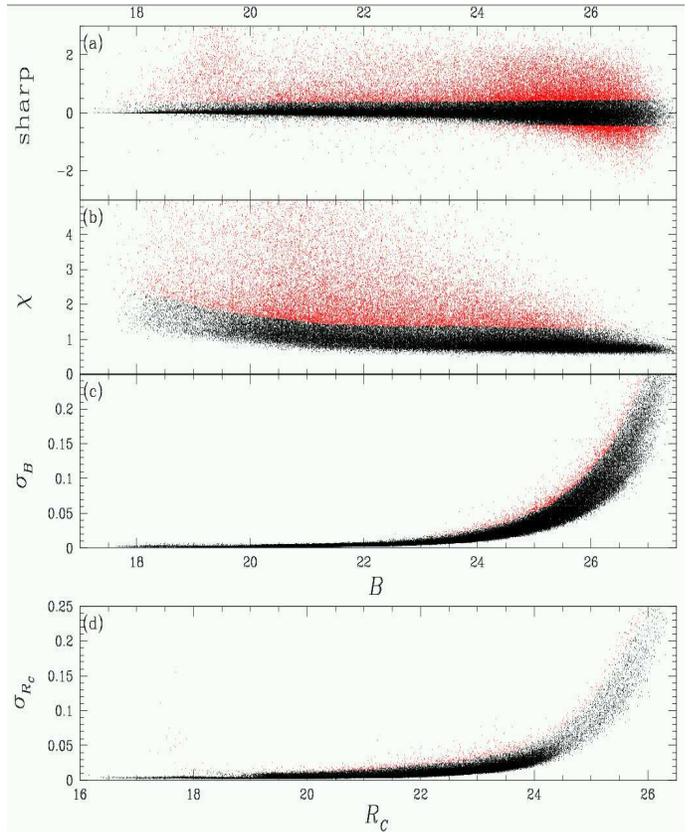} 
\caption{Selection criteria used to isolate FORS1@VLT stars for our MS
          subpopulation analysis. Panel (a) shows {\sf sharp} values
          versus $B$ magnitude, and panel (b) $\chi$ versus $B$.
          Panels (c) and (d) show the photometric errors as a function
          of the $B$ and Cousins-$R_C$ magnitudes respectively.  Only
          stars that passed the {\sf sharp} selection criterion (black
          in the first panel), are plotted in the subsequent panels;
          similarly, only stars that also survived the $\chi$
          selection are shown in the remaining two panels.}
\label{fig_fors_sel}
\end{figure}

%#####################################################################

\subsection{VLT: eight FORS1 fields}
\label{subsection2.FORS}

The VLT data set consists of eight partially overlapping FORS1 fields,
each with a pixel size of 200 mas and a field of view of
$6\farcm8\times6\farcm8$.  These fields (the blue rectangles in
Fig.~\ref{fig_fow}) probe the regions between $6\arcmin$ and
$25\arcmin$ from the center of \om.  The set of images consists of
$20\times1100$s exposures in $B$, and $20\times395$s in $R$, and are
the same images used by Sollima et al.\ (\cite{sollima07}).  There are
four images in each field (two per filter), except that the third and
fourth fields have four images per filter (see Fig.~\ref{fig_fow} for
field numbers).  This is the only data set that we reduced
specifically for the present work.  For this reason we give a more
detailed description of our reduction procedure.

We retrieved the data sets from the ESO archive; master-bias and
flat-field frames were constructed using standard IRAF routines.
Photometric reduction of the images was performed using P.\ Stetson's
DAOPHOT-ALLSTAR-ALLFRAME packages (Stetson \cite{stetson87},
\cite{stetson94}).  For each exposure we constructed a quadratic
spatially variable point-spread function (PSF) by using a Penny
function\footnote{A Penny function is the sum of a Gaussian and a
Lorentz function. In this case we used five free parameters:
half-width at half-maximum of the Penny function, in the $x$ and in
the $y$ coordinate; the fractional amplitude of the Gaussian function
at the peak of the stellar profile; the position angle of the tilted
elliptical Gaussian; and a tilt of the Lorentz function in a different
direction from the Gaussian.  The Lorentz function may be elongated
too, but its long axis is parallel to the $x$ or $y$ direction.}, and
for each individual exposure we chose---by visual inspection---the
best 100 (at least) isolated, bright, unsaturated stars that were
suitable for mapping the PSF variations all over the image.  We used
ALLFRAME on each individual field, keeping only stars measured in at
least four images.  The photometric zero points of each field were
registered to the instrumental magnitudes of the fourth field (the
less crowded of the two that have more exposures).  Finally,
photometric and astrometric calibration was performed using the
WFI@2.2$m$ astrometric-photometric catalog by Bellini et al.\
(\cite{bellini09}) as a reference.  As a result, we brought the FORS1
$R$ magnitudes to the Cousins-$R_C$ photometric system used by
WFI@2.2$m$.  Our final FORS1 catalog contains $\sim$133$\,$000
objects.

Since the innermost FORS1 field is seriously affected by crowding, we
did not use it in the present analysis.  Fig.~\ref{fig_fors_sel} plots
the {\sf sharp}, $\chi$, and $\sigma_B$ and $\sigma_{R_C}$ calculated
by ALLFRAME, as functions of stellar magnitude, for the stars in the
FORS1 catalog.  To choose the well-measured stars, we drew by eye the
cut-off boundaries in the quality parameters that retained objects
that were most likely to be well-measured stars.  Panel (a) shows {\sf
sharp} values versus $B$ magnitude.  Stars that passed the selection
criterion are shown in black.  Panel (b), which includes only stars
that passed the {\sf sharp} cut, shows $\chi$ values versus $B$.
Stars that also passed the $\chi$ criterion are in black.  In panel
(c) we plot the $\sigma_B$ values versus $B$, for the stars that
survived these two selections.  Again, the stars with good photometry
are shown in black.  Finally, in the last panel we plot $\sigma_{R_C}$
values versus $R_C$, for all the survivors, and we highlight in black
those that survived this selection too.  At the end of these selection
procedures, we are left with a catalog of $\sim$66$\,$500 stars.  We
note that while these selection criteria affect stars at different
magnitudes differently, they should not affect the ratio of stars on
the bMS and rMS, since at a given magnitude the two populations should
both have about the same photometric error, and the same probability
of making it into our catalog.

%#####################################################################

\subsection{WFI@2.2$m$}
\label{subsection2.WFI}

\begin{figure}[t!]
\centering
\includegraphics[width=9cm]{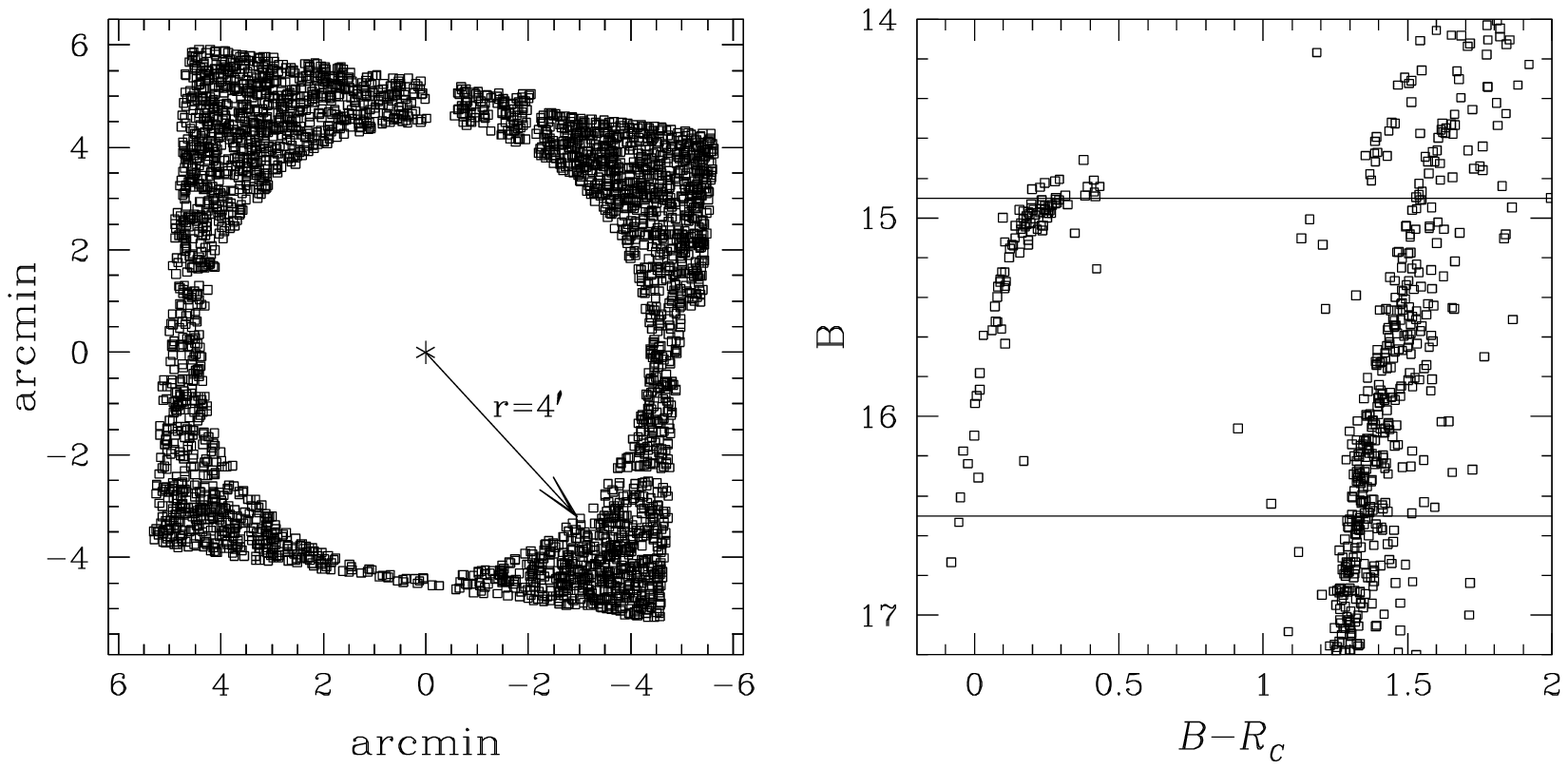}\\
\includegraphics[width=9cm]{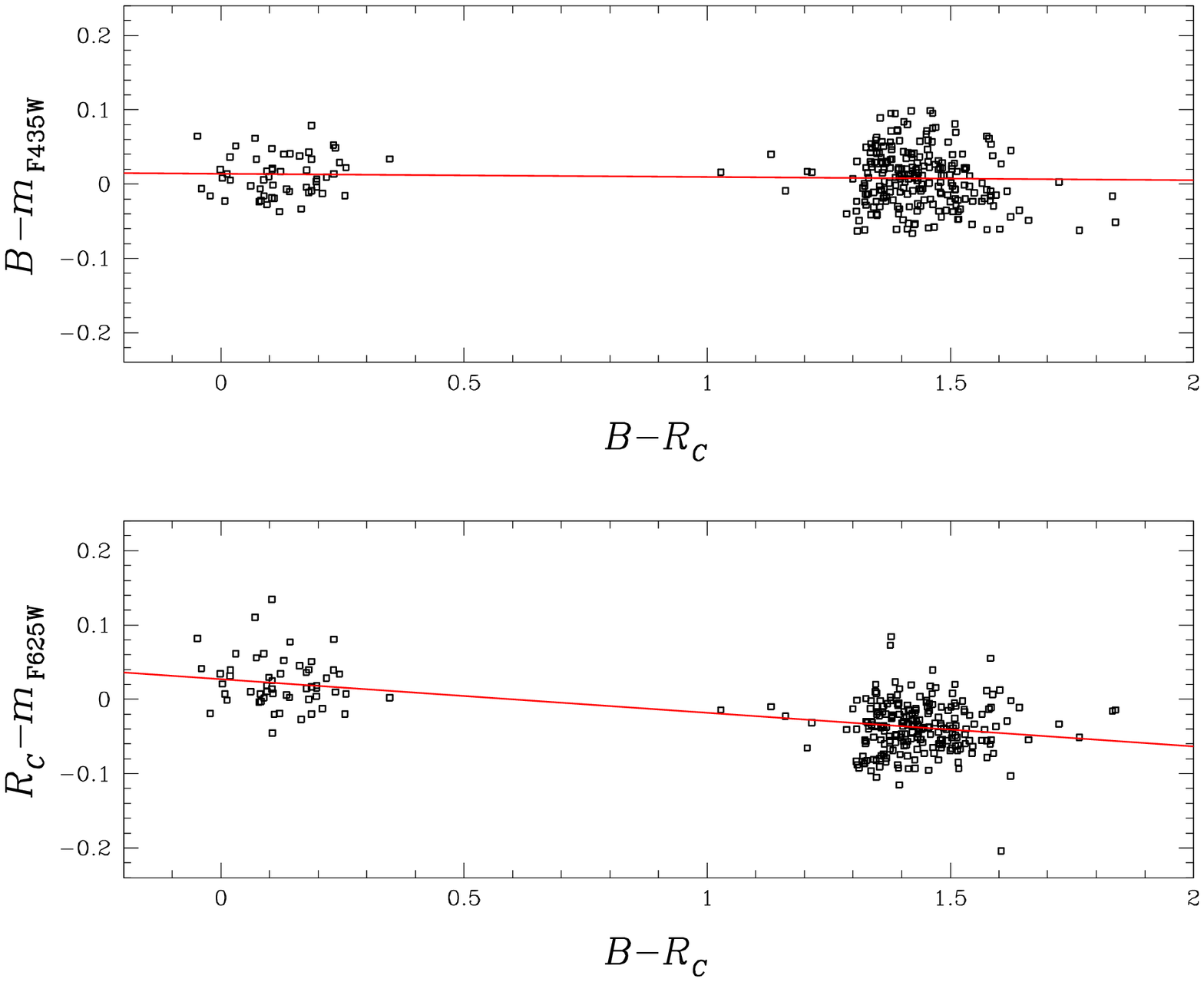}
\caption{(\textit{Top left:}) Selected stars in common between
    the \iacs\ and the WFI@2.2$m$ data sets. (\textit{Top right:})
    Horizontal lines mark the magnitude interval used to derive
    calibration equations. (\textit{Bottom panels:}) Calibration fits
    used to transform Vega-mag ACS/WFC $m_{\rm{F435W}}$ and
    $m_{\rm{F625W}}$ magnitudes into the WFI@2.2$m$ photometric
    system. See text for details.}
\label{fig:cal}
\end{figure}

This data set was collected at the 2.2$m$ ESO Telescope, with the WFI
camera, between 1999 and 2003.  The WFI@2.2$m$ camera is made up of a
mosaic of $4\times2$ chips, $2048\times4096$ pixels each, with a pixel
scale of 238 mas/pixel).  Thus, each WFI exposure covers
$\sim$34$\arcmin \times33\arcmin$.  The \om astrometric, photometric,
and proper-motion catalog based on this data set and presented in
Bellini et al.\ (\cite{bellini09}) is public, and contains several
wide-band ($U,B,V,R_C,I_C$) filters plus a narrow-band filter (658nm),
and covers an area of $\sim$33$\arcmin\times33\arcmin$ centered on the
cluster center.  We refer the reader to Bellini et al.\
(\cite{bellini09}) for a detailed discussion of the data-reduction and
calibration procedures.

Briefly, photometry and astrometry were extracted with the procedures
and codes described in Anderson et al.\ (\cite{anderson06a}).
Photometric measurements were corrected for ``sky concentration''
effects\footnote{Light contamination caused by internal reflections of
light in the optics, causing a redistribution of light in the focal
plane.}  and for differential reddening, as described in Manfroid \&
Selman (\cite{manfroid01}) and Bellini et al.\ (\cite{bellini09}).
Global star positions are measured to better than $\sim$45 mas in each
coordinate.  Photometric calibration in the $B,V,R_C,I_C$ bands is
based on a set of $\sim$3000 secondary standard stars in \om,
available on-line (Stetson \cite{stetson00}, \cite{stetson05}).  Color
equations were derived to transform our instrumental photometry into
the photometrically calibrated system using an iterative least-squares
linear fit.  Thanks to the four-year time-baseline, we were able to
successfully separate cluster members from field stars by means of the
local-transformation approach (Anderson et al.\ \cite{anderson06a}),
giving us proper motions more precise than $\sim$ 4 \masyr down to
$B\sim$20 mag, for $\sim$54$\,$000 stars.

%#####################################################################

\subsection{The astrometric and photometric reference frame}
\label{subsection2.ref.frame}

The large field of view of the WFI@2.2$m$ camera makes our WFI catalog
an ideal photometric and astrometric reference frame to which to refer
all the other observations, from different telescope-camera-filter
combinations.  For each catalog we made the tie-in by means of stars
that were in common with the reference catalog.  For positions we
derived a general six-parameter linear transformation to the
astrometric system of the WFI catalog.  For photometry we used as a
reference standard the $B$ and Cousins-$R_C$ magnitudes of the
WFI@2.2$m$ catalog, and transformed the magnitudes of each other
catalog to this standard.  For the $m_{\rm{F435W}}$ and
$m_{\rm{F625W}}$ magnitudes of the central mosaic of 3$\times$3
ACS/WFC fields, we used $\sim$3300 stars that had been observed in
common, located outside $4\arcmin$ from the cluster center to avoid
the most crowded regions in the WFI data set (top-left panel of
Fig.~\ref{fig:cal}).  We excluded from this sample saturated stars in
the WFI data set, keeping only the brighter ($14.9<B<16.5$) and well
measured ($\sigma_{B,R_C}<0.02$ mag) ones (top-right panel in
Fig.~\ref{fig:cal}).  The adopted calibration fits are shown in the
bottom panels of Fig.\ \ref{fig:cal}.  We did similarly for the FORS1
$B$ and $R$ magnitudes.

Calamida et al. (\cite{calamida05}) measured a differential reddening
of up to E$(B-V)\sim$0.14 in a region of $\sim$14$\arcmin \times $
14$\arcmin$ centered on \oma. This result has been questioned by
Villanova et al.\ (\cite{villanova07}); in their Figs.~1--6, the
sharpness of the SGB sequences suggests that the existence of any
serious differential reddening is very unlikely.  But in any case, a
proper radial-distribution analysis needs correction even for a
differential reddening that is of the order of few hundredths of a
magnitude.  Our corrections for differential reddening followed the
method outlined by Sarajedini et al.\ (\cite{sarajedini07}), which
uses the displacements of individual stars from a fiducial sequence to
derive a reddening map.

The outer ACS/WFC field at $\sim$17$\arcmin$ from the cluster center
and the WFPC2 field at $\sim$7$\arcmin$ provide stellar photometry in
the ${\rm F606W}$ and ${\rm F814W}$ bands.  For the ACS field we have
overlap with the WFI catalog, which allows us to calibrate the
photometry, but the stars available are all on the main sequence above
$m_{\rm F606W}=21$, so they have a very narrow range in color, and we
cannot empirically determine the color term in the calibration.  For
the WFPC2 field, in addition to the problem of the limited color
baseline, the WFI photometry in this inner field is of low quality on
account of ground-based crowding.  For these reasons, we decided to
not transform the photometry of these two fields into the photometric
reference system of \wfi, but dealt with them in the {\it HST}
Vega-mag flight system.

%#####################################################################

\begin{figure}[t!]
\centering
\includegraphics[width=9.0cm,height=9.0cm]{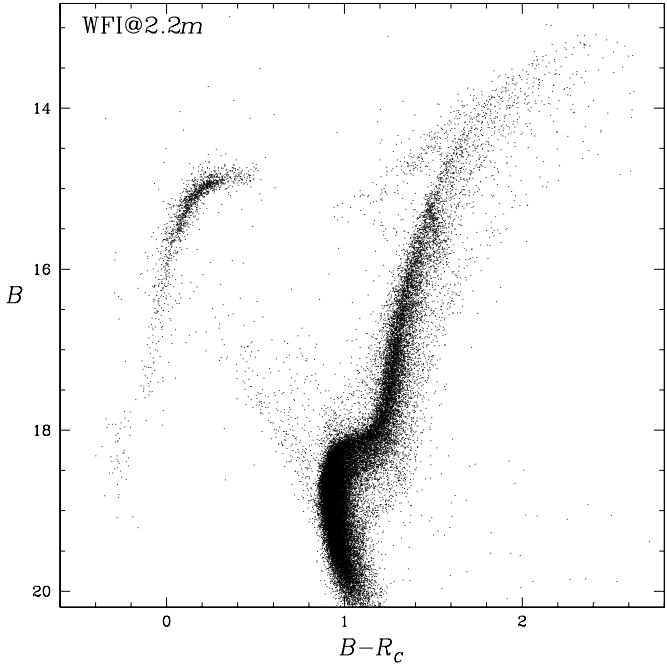}
\caption{$B$ vs. $B-R_C$ CMD of WFI@2.2$m$ stars, after calibration
         and proper-motion selection (see Bellini et
         al. \cite{bellini09}).}
\label{fig:wfi_cmd}
\end{figure}

\subsection{The deep color-magnitude diagrams}
\label{subsec_cmd}

Our proper-motion-selected \wfi\ $B$ vs.\ $B-R_C$ CMD is shown in
Fig.~\ref{fig:wfi_cmd}.  All the main features of the cluster are
clearly visible, except for the split MS, since the WFI data go down
only a magnitude or so below the turnoff.  The CMDs of the other data
sets that we analyzed are presented in Fig.~\ref{fig:cmds_others},
where the top-left panel refers to the eight FORS1@VLT fields, the
middle-left panel to the proper-motion-selected CMD of the external
ACS/WFC, the bottom-left panel the CMD from the WFPC2 field, and the
right panel of Fig.~\ref{fig:cmds_others} the CMD of the inner
$3\times3$ ACS/WFC mosaic.  It is clear that the MS population can be
studied in all but the WFI CMD, and the RGB population can be studied
in the WFI and inner ACS data sets.

Now that we have a comprehensive sample of \om stars, both for the
bright stars and for the faint ones, covering the central region all
the way out to $\sim$25$\arcmin$, we can define robust selection
criteria for the subpopulations to track how the population fractions
vary with radius.

%#####################################################################

\subsection{The \emph{angular} radial distance: $r^*$}
\label{subsec:r*}

Since $\omega$ Cen is elongated in the plane of the sky, it does not
make sense to analyze its radial profile via circular annuli.  We
therefore decided to extract radial bins in the following way.  We
adopted the position angle (P.A.) of $100^\circ$ for the major-axis
(van de Ven et al.\ \cite{vdv06}), and an average ellipticity of 0.17
Geyer et al. \cite{geyer83}).  To define the bins of the radial
distribution we adopted elliptical annuli, whose major axes are
aligned with the \om major axis, and stars were extracted accordingly
(see Fig.~\ref{fig_fow}).  To indicate the angular radial distance
from the cluster center, we used the equivalent radius \re, defined as
the radius of the circle with the same area as the corresponding
ellipse (i.e., the geometrical mean of the semi-major and semi-minor
axes).  Each of the small fields (the outer ACS field and the WFPC2
field), we considered as a single radial bin.

%#############################################################

\begin{figure*}[ht!]
\centering
\includegraphics[height=20.5cm]{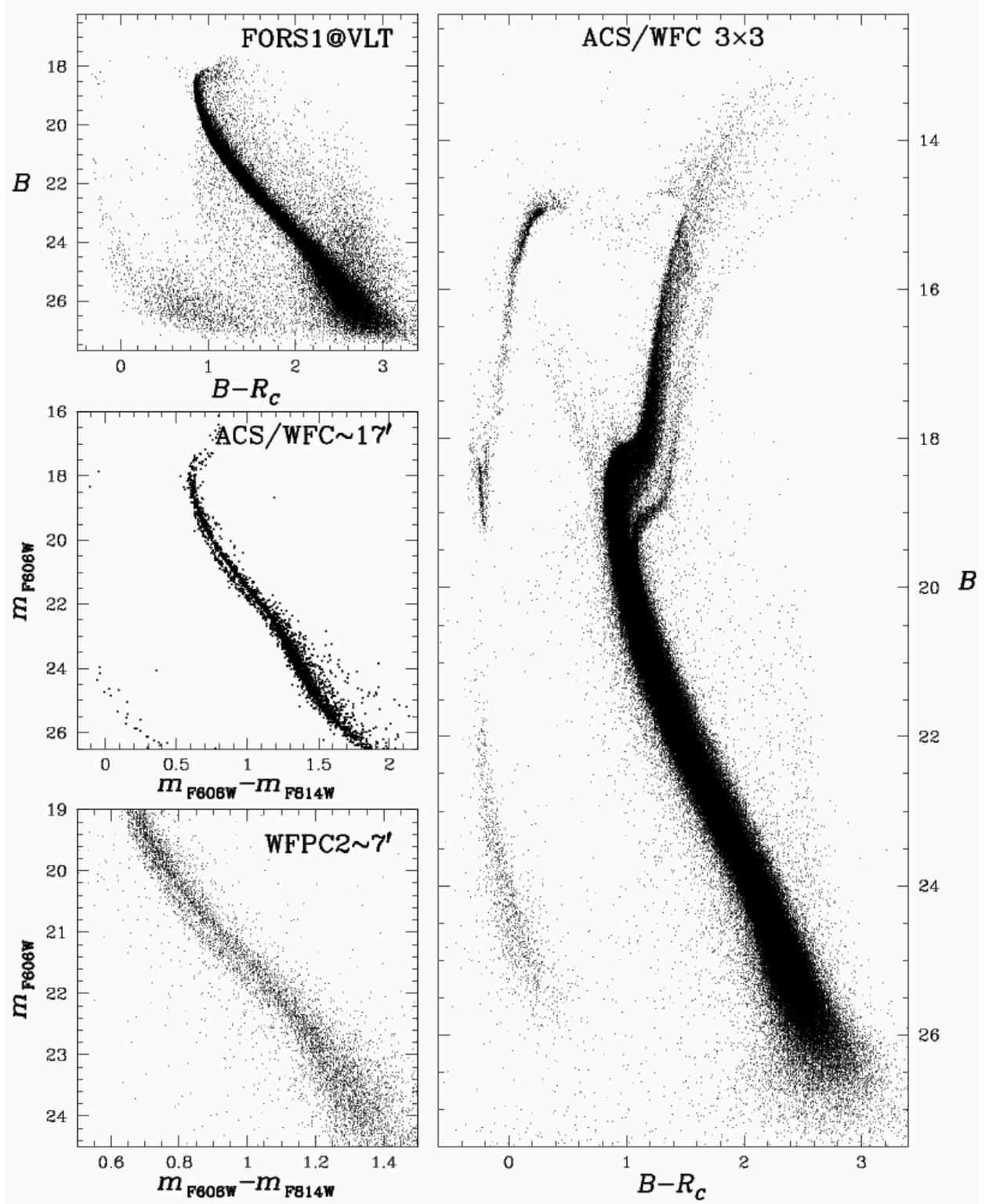} 
\caption{ (\textit{Top left}): CMD from the eight FORS1@VLT fields.
                We can measure stars from the bottom of the RGB down to
                $B\sim$27.5 mag.  (\textit{Middle left}):
                proper-motion-selected CMD from the outer ACS/WFC
                field.  (\textit{Bottom left}): CMD from the WFPC2
                images located $\sim$7$\arcmin$ south of the cluster
                center.  (\textit{Right panel}): CMD of the inner
                $3\times3$ ACS/WFC fields.  In the top left and the
                right-hand CMDs, the bMS and rMS fail to show
                separately only because the profusion of points
                blackens their whole region.}
\label{fig:cmds_others}
\end{figure*}

%%%%%%%%%%%%%%%%%%%%%%%%%%%%%%%%%%%%%%%%%%%%%%%%%%%%%%%%%%%%%

\section{MS subpopulations}
\label{sec:mss}

Our goal in putting together these varied catalogs is to quantify the
differences in the radial profiles of the various subpopulations of
$\omega$ Cen.  One way to do this would be to measure the surface
density profile for each group and compare them directly, but this
would require accurate completeness corrections and careful attention
to magnitude bins.  Since our interest, however, is simply to
determine how the populations vary with respect to each other, we need
only measure the {\it ratio} of the populations as a function of
radius.  This ratio should be independent of completeness corrections
and the details of the magnitude bins used, since the bMS and rMS
differ only slightly in color and are observed over the same magnitude
range.

Our analysis of the \nbr\ ratio is based mostly on the data sets from
the inner \iacs\ and FORS1@VLT, which allow us to map the ratio of
bMS/rMS from the cluster center out to $\sim$25$\arcmin$, once the
photometry and astrometry have been brought into the same reference
system.  The other two fields, each of which covers only a small
region, provide only one point each in our analysis of \nbr\ versus
radius.  Moreover, since we were not able to bring $m_{\rm F606W}$ and
$m_{\rm F814W}$ photometry of the outer ACS and the WFPC2 field into
the WFI $B$ and $R_C$ photometric system, we kept the WFPC2 and the
outer ACS/WFC data sets in their native photometric system, and used
them only for a further (though important) confirmation of the radial
gradient found with the FORS1 and inner ACS/WFC data sets.

%#####################################################################

\subsection{Straightened main sequences}
\label{subsec:MS1}

\begin{figure}[t!]
\centering
\includegraphics[width=9cm]{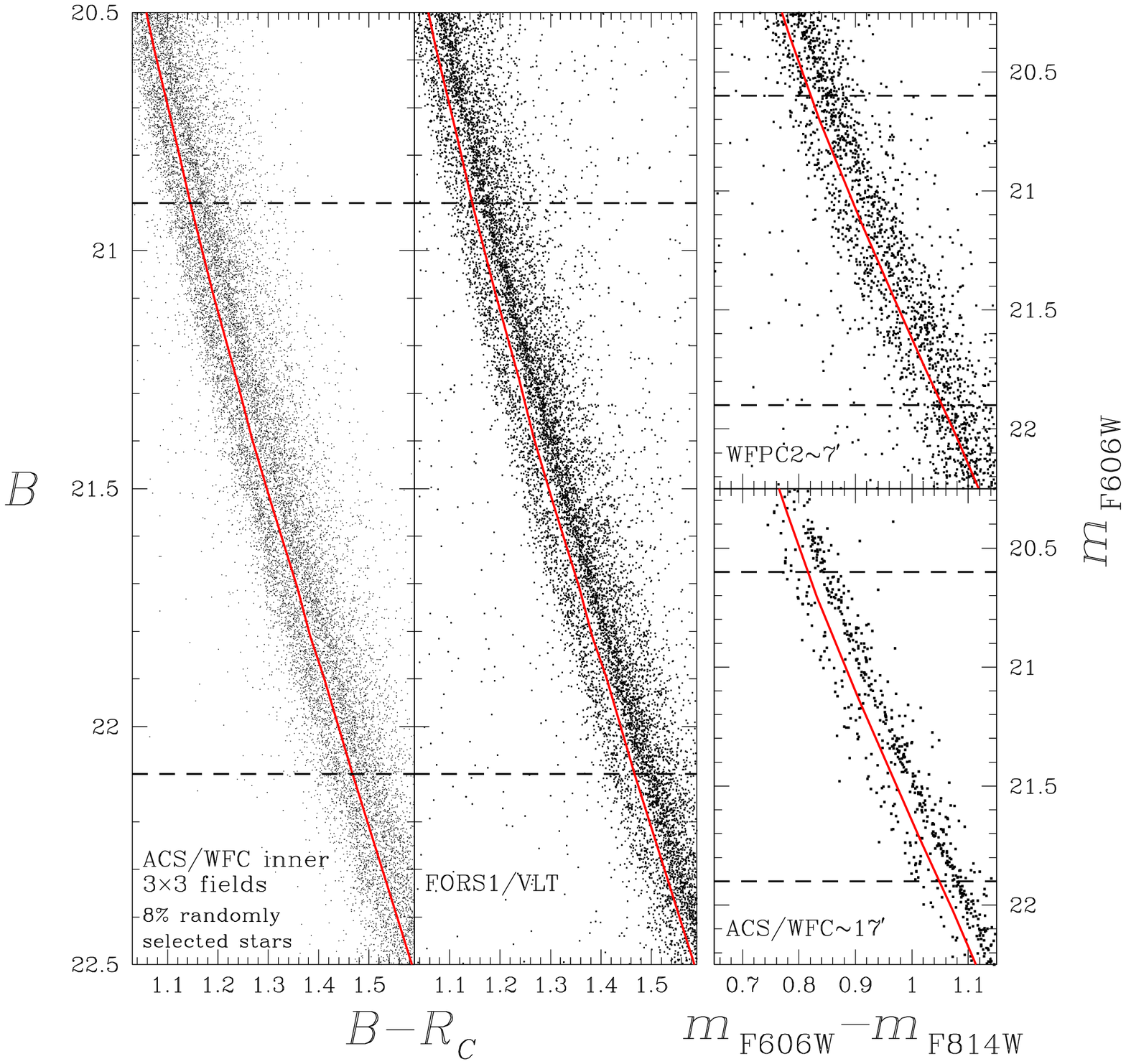}
\caption{The left panel shows a randomly selected 8\% of the stars in
          the CMD of the inner $\sim$10$\arcmin\times10\arcmin$
          ACS/WFC images, in the region of the MS where the two
          branches are most separated in color.  The middle panel
          shows the CMD of the FORS1@VLT fields.  The right panels
          show the outer ACS/WFC field (bottom) and WFPC2 field (top).
          The MS duality is clearly detected in all diagrams (see also
          Fig.~\ref{fig:cmd_fig3c}).  The dashed horizontal lines mark
          the selected magnitude range for the definition of the bMS
          and rMS samples used in the derivation of their radial
          profiles.  The fiducial lines (drawn by hand) that were used
          to straighten and separate the sequences are also plotted
          (in red in the color version).}
\label{fig:cmd_fig3b}
\end{figure}

\begin{figure}[t!]
\centering
\includegraphics[width=9cm]{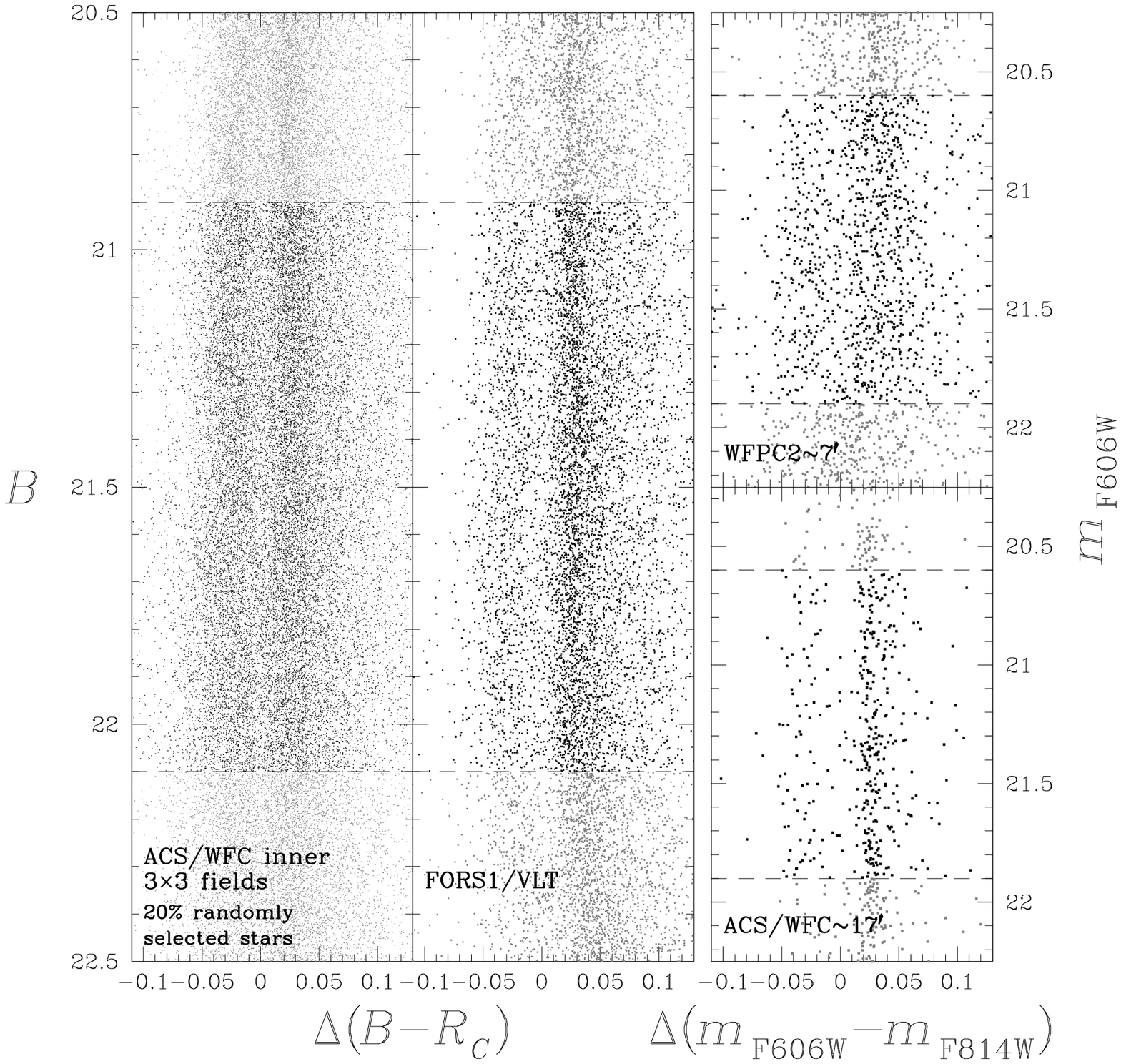}
\caption{Same as Fig.~\ref{fig:cmd_fig3b}, but after subtraction, from
          the color of each star, of the color of the fiducial line at
          the same luminosity.  In the left panel we show a randomly
          selected 20\% of the stars from the ACS/WFC central-mosaic
          data (rather than the previous 8\%, since the color-scale is
          now less compressed).}
\label{fig:cmd_fig3c}
\end{figure}

In order to analyze the color distribution of the stars along the MS
in a more convenient coordinate system, we adopted a technique
previously used with success in \om~ (Anderson \cite{anderson97},
\cite{anderson02}), and in other works (Sollima et al.\
\cite{sollima07}; Villanova et al.\ \cite{villanova07}; Piotto et al.\
\cite{piotto07}; Milone et al.\ \cite{milone08}, Anderson et al.\
\cite{anderson09a}).

We defined fiducial lines in the CMDs (drawn by hand), such as to be
equidistant from the ridge lines of the bMS and rMS stars.  We avoided
choosing the ridge line of either sequence as our fiducial line,
because we wanted a system in which both the sequences are as parallel
and as rectified as possible.  We used different fiducial lines for
the $B$, $B-R_C$ CMDs of the inner ACS/WFC and the FORS1 data sets and
for the ($m_{\rm F606W}$, $m_{\rm F606W}-m_{\rm F814W}$) CMDs of the
WFPC2 and outer ACS/WFC data sets.  In this way, we were sure to
straighten the MSs in the same consistent way for the two different
sets of filters.  Then we subtracted from the color of each star the
color of the fiducial line at the same luminosity as the star.

In Fig.~\ref{fig:cmd_fig3b} we show the CMDs in the \om MS region for
the central mosaic of ACS/WFC data (left panel), the FORS1@VLT (middle
panel), and the WFPC2 $\sim$7$\arcmin$ field and the ACS/WFC field at
$\sim$17$\arcmin$ (right panels).  In the case of the central ACS/WFC
data, we plotted only a randomly chosen 8\% of the stars, in order to
show the two sequences clearly.  In all the CMDs the MS splitting is
clearly visible.  For the inner ACS/WFC and FORS1 data sets we
restricted our MS analysis to the magnitude range $20.9\leq B\leq22.1$
(dashed lines in Fig.~\ref{fig:cmd_fig3b}), the interval in which the
two MSs are most separated in color and are parallel.  For the same
reasons we analyzed stars in the magnitude range $20.6\leq
m_{\rm{F606W}}\leq21.9$ for the WFPC2 and the outer ACS/WFC data sets.
The bright limit also avoids the saturated stars in the deep WFC
exposures.  The adopted fiducial lines are again plotted (in red in
the color version of the paper).

In Fig.~\ref{fig:cmd_fig3c} we show straightened CMDs for the same
data sets shown in Fig.~\ref{fig:cmd_fig3b}, with the only difference
being that we now plot a 20\% randomly generated sample of stars for
the inner ACS/WFC data set, since the expanded color baseline allows
more points to be seen.  It is worth noting that even a simple
inspection shows the \nbr\ ratio clearly decreasing as we go from the
central cluster regions to the outer ones.  It is also clear that the
spread in the bMS is somewhat greater than that of the rMS.

Finally, note that we call the color deviation of a star from the
fiducial line $\Delta(B-R_C)$.  We shall use this notation frequently
in what follows.

Our aim in selecting the best-measured stars in the previous sections
was so that we would be able to assign the stars to the different
populations as accurately as possible.  Similarly, as much as possible
we transformed our photometry into the same system, so that our
population selections throughout the cluster would be as consistent as
possible.

Even with these careful steps, however, it is still difficult to
ensure that we are selecting stars of the same population in the inner
parts of the cluster as in the outer parts.  Even if we had
observations with the same detector at all radii, the greater crowding
at the center would increase the errors there.  On the other hand, our
use of ground-based images for the outer fields actually makes those
fields even {\it more} vulnerable to crowding effects.

Another complication comes from main-sequence binaries, which at the
distance of a globular cluster are unresolved.  Relaxation, causing
mass segregation, will concentrate them to the cluster center and
cause a redward distortion of the main sequence there.

Moreover, in the lower-density outer regions of the cluster we can get
the same statistical significance only by using larger areas, with an
increased vulnerability to inclusion of field stars.  Finally, the red
side of the main sequence is contaminated by the anomalous metal-rich
population (hereafter MS-a), which is clearly connected with RGB-a.
Even if these stars include only $\sim$5\% of the total cluster
members (Lee et al.\ \cite{lee99}; Pancino et al.\ \cite{pancino00};
Sollima et al.\ \cite{sollima05a}; Villanova et al.\
\cite{villanova07}), they are an additional source of pollution for
rMS stars---against which we now take specific precautions.

%##############################################################

\subsection{Dual-Gaussian fitting}
\label{subsec:gauss}

\begin{figure}[t!]
\centering
\includegraphics[width=8.7cm]{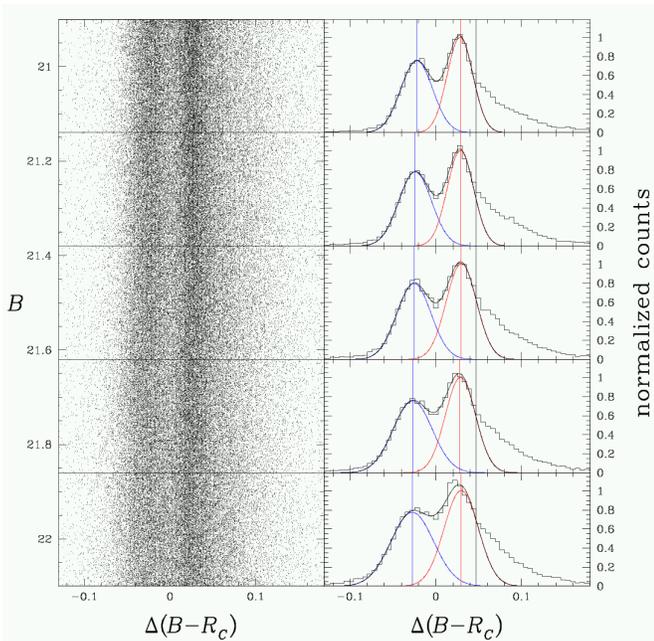}
\caption{(\textit{Left panel}): $B$ versus $\Delta(B-R_C)$ diagram for
               selected stars in our data set from the \iacs, divided
               into five magnitude intervals.  We now plot {\it all}
               the stars in this sample, not just a randomly selected
               subset.  (\textit{Right panels}): $\Delta(B-R_C)$
               histograms with the Gaussian best fits. See text for
               details.}
\label{fig:Y3bins}
\end{figure}

There is no way of dealing with the above issues perfectly, but we did
our best to make our measurements as insensitive to them as possible.
To this end, we measured the bMS and rMS fractions by simultaneously
fitting the straightened color distributions with two Gaussians, and
taking the area under each Gaussian as our estimate of the number of
stars in each population.  By keeping the width of each Gaussian an
adjustable parameter, we allowed in a natural way for the fact that
the photometric scatter differs from one radius and data set to
another.

While the dual Gaussians provide a natural way of measuring the two
populations in data sets that have different color baselines and
different photometric errors, there is one serious complication.  As
we have indicated, there is an unresolved, broad population of stars
redward of the rMS that consists of blends, binaries, and members of
the MS-a branch.  Since it is unclear what relation this mixed
population has with the two populations that we are studying, we
wanted to exclude it from the analysis as much as possible.  We did so
by cutting off the reddest part of the color range, and confining our
fitting to the color range that is least disturbed by the contaminated
red tail.

In order to choose the red cutoff as well as possible, we gathered
together all of the stars in each data set. Below we will describe for
simplicity only the case of the central 3$\times$3 mosaic of ACS
images in $B$ and $R_C$. The procedure followed is, however, the same
for the other data sets.

Within this data set we chose the MS stars that were in the magnitude
range $20.9\leq B \leq22.1$ (within which the two MSs are almost
parallel and are maximally separated in color) and in the color range
$-0.25\leq \Delta(B-R_C)\leq 0.25$ mag.  We emphasize that this
ensemble of the data set, within which we will later see a
considerable gradient in the relative numbers of bMS and rMS stars,
will not be used to derive population results in the case of the inner
ACS/WFC data set, but only to choose the red cutoff.  We divided these
stars into five magnitude intervals, because the observational errors,
which increase the spread of the sequences, depend on magnitude.
Next, we plotted histograms of the $\Delta(B-R_C)$ distribution within
each magnitude interval, using a bin size of 0.006 mag.  This size is
$\sim$1/4 of the typical photometric error in color; it makes a good
compromise between a fine enough color resolution, on the one hand,
and adequate statistics, on the other hand.

The actual choice of the red cutoff is a two-tiered procedure.  We
must first develop a procedure for the fitting of dual Gaussians to a
set of bins that has a red cutoff; then we must decide on a value of
$N_{\rm red}$, the number of bins that we include on the redder side
of the red Gaussian.

\begin{figure*}[t!]
\centering
\includegraphics[width=\textwidth]{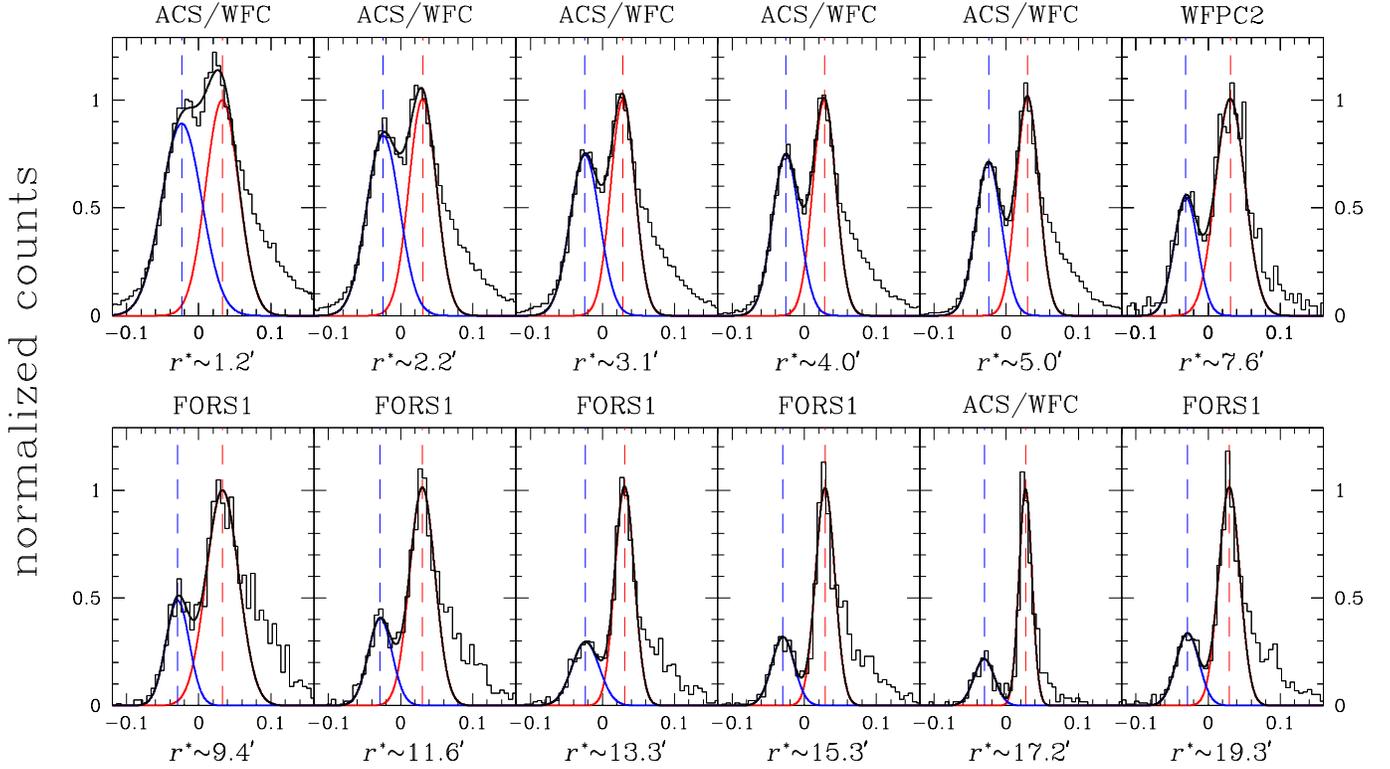}
\caption{Dual-Gaussians fits.  As in Fig.\ \ref{fig:Y3bins}, the
     Gaussian fits to the bMS and rMS are in blue and red
     respectively, and their sum in black.  The vertical dashed lines
     mark the centers of the individual Gaussians.  The individual
     panels are arranged in order of effective radius.  (Note that
     {\it all} our fields are shown here, in radial order, so that the
     WFPC2 field follows the inner ACS fields, and the outer ACS field
     falls between two of the FORS1 fields.)}
\label{fig_dualGauss_fits}
\end{figure*}

Although from a mere inspection of the histograms it is clear where,
approximately, the peak of the red Gaussian should lie, the narrowness
of the bins leaves it uncertain in which particular bin the peak of
the red Gaussian will actually fall.  Since the red cutoff, $N_{\rm
red}$, is defined as being counted from that bin, we had to resort to
an iterative procedure to locate the cutoff for a given value of
$N_{\rm red}$.  We began by choosing a cutoff safely to the red of
where we guessed that the cutoff would actually fall, and then using
that cutoff in a first try at fitting the dual Gaussians.  The
iteration then consisted of placing the cutoff just beyond $N_{\rm
red}$ bins on the red side of the peak of the red Gaussian and fitting
again; this new fit might cause the red peak to move to a different
bin.  When the red peak stays in the same bin, the iteration has
converged; this happened after very few iterations.

We assumed trial values of $N_{\rm red}$ from 2 to 5, and for each of
those values we iteratively computed the Gaussian parameters for each
of the five magnitude intervals.  We chose as the best value for
$N_{\rm red}$ the one for which the five values of $N_{\rm bMS}/N_{\rm
rMS}$ were the most consistent.  This value turned out to be $N_{\rm
red}=3$.  With this choice made, we then moved on to fit dual
Gaussians to each of our detailed data sets.

Fig.~\ref{fig:Y3bins} shows the results of this procedure.  In the
left panel we show our selected stars in the $B$ versus
$\Delta(B-R_C)$ diagram---all of the stars this time, rather than a
random selection of a fraction of them.  The horizontal lines
delineate our five magnitude intervals.  On the right we show the
final $\Delta(B-R_C)$ histogram for each magnitude interval, and the
dual-Gaussian fit to it.  The individual Gaussians are shown in blue
and red, respectively, and the black curve is their sum.  The vertical
blue and red lines are the centers of the respective Gaussians, and
the vertical black line shows the red cutoff.  Note that we do not
show the vertical boundaries between the bins of a histogram, because
on this scale they would be too close to each other.  Nor do we show
the Poisson errors of the counts in the bins, because they are small
and would obscure the bin values themselves; the size of the errors is
amply clear from the smoothness of the values in neighboring bins.
The counts in the histograms are normalized so as to make the height
of the red Gaussian equal to unity.

\begin{figure*}[t!]
\centering
\includegraphics[width=\textwidth]{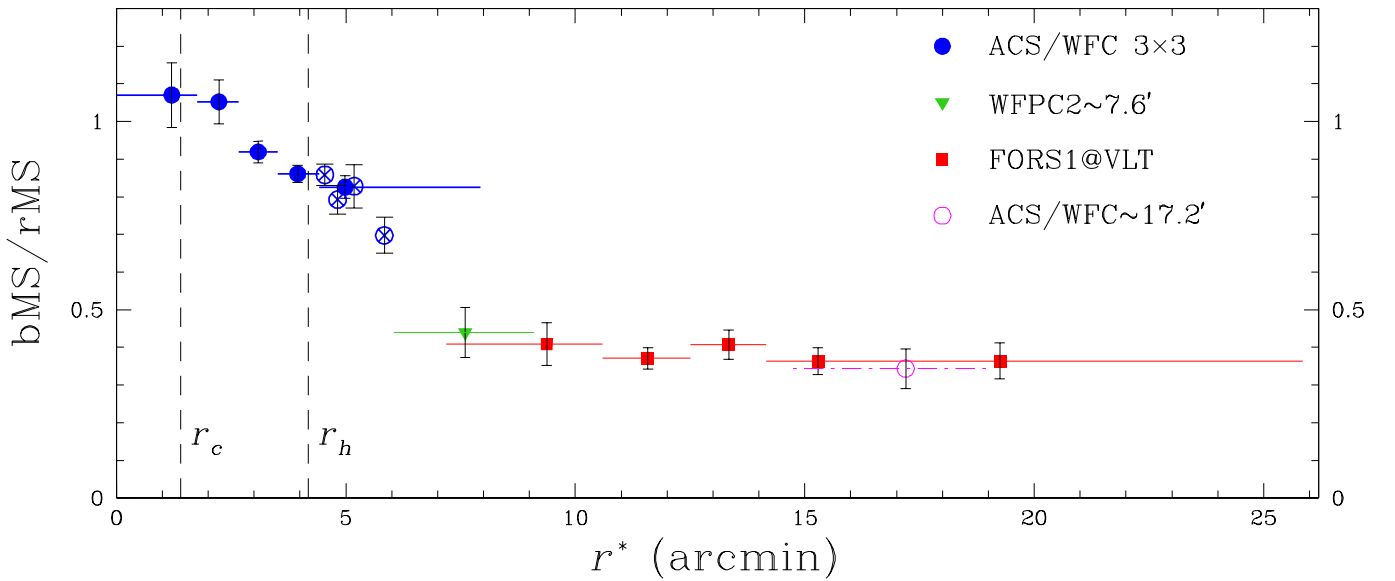}
\caption{\nbr\ ratio versus equivalent radius \re.  Different colors
    and symbols refer to different data sets.  Dashed vertical lines
    mark the core radius and the half-mass radius.  Error
    bars were calculated from the residuals of values in individual
    subdivisions (quadrants for the inner ACS/WFC mosaic, magnitude
    intervals in each outer field).  To improve the radial resolution
    for the outermost annulus of the inner ACS/WFC mosaic, we also
    divided it into four sub-annuli (crossed open circles).  See text
    for a fuller explanation.}
\label{fig_gradient}
\end{figure*}

\begin{table*}[t!]
\tiny{
\centering
\caption{Dual-Gaussian fitting results.  For each data set (first
          column) we give in Cols.\ 2--4 the radial extent (minimum,
          median, and maximum \re).  In Cols.\ 5--8 are the sigmas of
          the Gaussians that best fit the bMS and rMS color
          distributions, with errors. In the next two columns are the
          \nbr\ ratio and its error.  The next-to-last column gives
          the color difference between the two Gaussian peaks, and the
          final column identifies the color baseline of the data set.}
\begin{tabular}{cccccccccccc}
\hline
\hline
&&&&&&&\\
$\!\!\!$data set &$\!\!\!$$r^*$ min $\!\!\!$&$\!\!\!$ $r^*$ median
                       $\!\!\!$&$\!\!\!$ $r^*$max
                       $\!\!\!$&$\!\!\!$$\sigma_{\rm{bMS}}$
                       $\!\!\!$&$\!\!\!$
                       rms($\sigma_{\rm{bMS}}$) $\!\!\!$&$\!\!\!$ $\sigma_{\rm{rMS}}$ $\!\!\!$&$\!\!\!$
                       rms($\sigma_{\rm{rMS}}$) $\!\!\!$&$\!\!\!$ \nbr &$\sigma$(\nbr)$\!\!\!$&$\!\!\!$
                       $({\rm rMS}_{\rm cen}-{\rm bMS}_{\rm cen})$ $\!\!\!$$\!\!\!$&$\!\!\!$$\!\!\!$ color\\
                       & ($\arcmin$) & ($\arcmin$) & ($\arcmin$) &
                       $\Delta$color & $\Delta$color & $\Delta$color &
                       $\Delta$color &&  &$\Delta$color &\\
&&&&&&&\\
\hline
&&&&&&&\\
ACS/WFC        & 0.00 & 1.21 & 1.76 & 0.027 & 0.0020 & 0.023 & 0.0014 & 1.07 & 0.09& 0.056 & $B-R_C$\\
   (3$\times$3) & 1.76 & 2.24 & 2.66 & 0.023 & 0.0018 & 0.018 & 0.0011 & 1.05 & 0.06& 0.055 & $B-R_C$\\
                & 2.66 & 3.09 & 3.51 & 0.020 & 0.0012 & 0.017 & 0.0008 & 0.92 & 0.03& 0.053 & $B-R_C$\\
                & 3.51 & 3.95 & 4.42 & 0.018 & 0.0010 & 0.016 & 0.0007 & 0.86 & 0.02& 0.054 & $B-R_C$\\
                & 4.42 & 4.98 & 7.93 & 0.018 & 0.0011 & 0.015 & 0.0007 & 0.82 & 0.03& 0.054 & $B-R_C$\\
&&&&&&&\\
    subdivision & 4.42 & 4.54 & 4.67 & 0.019 & 0.0012 & 0.015 & 0.0011 & 0.86 & 0.03&	    &	   \\
    of last bin & 4.67 & 4.82 & 4.98 & 0.018 & 0.0013 & 0.016 & 0.0011 & 0.79 & 0.04&	    &	   \\
                & 4.98 & 5.18 & 5.44 & 0.018 & 0.0013 & 0.015 & 0.0010 & 0.83 & 0.06&	    &	   \\
                & 5.44 & 5.84 & 7.93 & 0.018 & 0.0013 & 0.015 & 0.0010 & 0.70 & 0.05&	    &	   \\
&&&&&&&\\ 
WFPC2          & 6.04 & 7.57 & 9.10 & 0.017 & 0.0010 & 0.020 & 0.0020 & 0.42 & 0.07& 0.061 &$\!\!\!\!\!\!\!\!\!\!\!\!\!\!m_{\rm F606W}\!-\!m_{\rm F814W}$\\
&&&&&&&\\ 
FORS1          & 7.18 & 9.38 & 10.60& 0.017 & 0.0023 & 0.020 & 0.0021 & 0.41 & 0.06& 0.062 & $B-R_C$\\
                & 10.60& 11.58& 12.51& 0.017 & 0.0019 & 0.017 & 0.0013 & 0.37 & 0.03& 0.058 & $B-R_C$\\
                & 12.51& 13.34& 14.16& 0.019 & 0.0018 & 0.014 & 0.0009 & 0.41 & 0.04& 0.054 & $B-R_C$\\
                & 14.16&15.29 & 16.75& 0.016 & 0.0022 & 0.014 & 0.0009 & 0.36 & 0.04& 0.059 & $B-R_C$\\
                & 16.75&19.25 & 26.19& 0.016 & 0.0020 & 0.014 & 0.0010 & 0.36 & 0.05& 0.058 & $B-R_C$\\
&&&&&&&\\ 
ACS/WFC        & 14.68& 17.21& 19.69& 0.014 & 0.0020 & 0.009 & 0.0020 & 0.34 & 0.05& 0.057 &$\!\!\!\!\!\!\!\!\!\!\!\!\!\!m_{\rm F606W}\!-\!m_{\rm F814W}$\\
&&&&&&&\\ 
\hline
\end{tabular}
\label{tab:MS}
}
\end{table*}

%####################################################################

\subsection{The Radial Gradient of \nbr}
\label{subsec:method1}

Having chosen the position of the red cutoff, we were able to perform
dual-Gaussian fitting on each of our data
sets. Figure~\ref{fig_dualGauss_fits} shows our fits.  We divided the
inner ACS/WFC mosaic and the outer FORS1@VLT data sets into five
radial intervals for each.  The intervals were chosen in such a way as
to have the same number of selected stars in each of them, so that the
statistical sampling errors will be uniform.  (The reader should note
that Fig.\ \ref{fig_dualGauss_fits} shows {\it all} of our fields, in
radial order, so that the WFPC2 field follows the inner ACS fields,
and the outermost ACS field falls between two of the FORS1 fields.)

Figure \ref{fig_gradient} shows our results for the radial variation
of the bMS to rMS ratio, for the five radial parts of the inner ACS
mosaic, the five radial intervals of our FORS1 fields, the WFPC2
field, and the outer ACS field.  Symbols of a different shape
distinguish the various types of field.  The outermost radial interval
of the ACS/WFC mosaic is a special case, however, since it consists
largely of the four corners of the mosaic, and it spans a larger
radial extension. To better map the bMS/rMS distribution in this
radial interval, we decided to further split it into four sub-annuli.
In this way we increase the radial resolution, but pay the price of
larger sampling errors.  We have therefore plotted the outermost
radial interval of the inner ACS/WFC mosaic twice, once as a whole
annulus, and once as four sub-annuli (marked as crossed open circles
in Fig.\ \ref{fig_gradient}).

Our choice of using ellipses with fixed ellipticity and position angle
to extract radial bins could have introduced some systematics in our
derived \nbr\ ratios. To address this issue, we recalculated the \nbr\
ratios by extracting radial bins using simple circles, and we found no
significant differences between the two radial binning methods.

Estimating the errors of our points required special attention.  First
we took the Poisson errors of the numbers of stars, and used them to
generate Poisson errors for the values of \nbr.  These, however, are
only a lower bound for the true error, which has additional
contributions that are impossible to estimate directly; they come from
blends, binaries, etc.  To estimate the true errors empirically, for
each value of \nbr\ we subdivided the sample of stars that had been
used.  In the inner ACS/WFC mosaic the subsamples were the quadrants
shown in Fig.\ \ref{fig_fow}, while for each of the outer fields,
where we do not have symmetric azimuthal coverage, we divided the
sample into magnitude intervals, four for each FORS1@VLT field and
three each for the WFPC2 field and the outer ACS/WFC field.

We treated each set of subsamples as follows: Within each subsample we
performed a dual-Gaussian fit, and derived from it the value of \nbr.
We weighted each subsample according to the number of stars in it, and
took a weighted mean of the four (or three) values of \nbr, to verify
that this mean was equal, within acceptable round-off errors, to the
value that we had found for the whole sample.  (It was, within a per
cent or two in nearly every case.)  Finally we derived an error for
the sample, from the residuals of the individual \nbr\ values from
their mean, using the same weights as we had used for the mean.  These
are the error bars that are shown in Figure \ref{fig_gradient}.  These
errors are indeed larger than the Poisson errors, but only by about
10\%.  We must note, however, that in addition to the random error
represented by the error bars, it is likely that there is still some
systematic error in our values of \nbr, due to the effects of
blends and binaries.  On the one hand, blends have the same
photometric effect as true binaries; they tend to move bMS stars into
the rMS region, while many of the rMS stars that are similarly
affected are eliminated by our red cut-off.  This effect tends to
reduce our observed value of \nbr.  It is less easy to predict,
however, how such effects increase toward the cluster center.  Blends,
on the one hand, increase because of the greater crowding.  Binaries,
on the other hand, increase because their greater mass gives them a
greater central concentration.  To repeat, the result has been that
our values of \nbr\ are somewhat depressed toward the cluster center,
so that the gradient of \nbr\ that we report is probably a little
lower than the real one.

Fig.~\ref{fig_gradient} clearly shows a strong radial trend in the
ratio of bMS to rMS stars, with the bMS stars more centrally
concentrated than the rMS stars.  The most metal-rich population,
MS-a, is too sparse, and also too hopelessly mixed with the red edge
of the rMS, to allow any reliable measurement of its radial
distribution, but in the next section we will examine the distribution
of its progeny, RGB-a.  Table \ref{tab:MS} summarizes our results.
The first column identifies the data set.  Columns 2--4 give, for the
inner ACS/WFC 3x3 mosaic, the minimum, median, and maximum radius of
the central circle or the annulus, while for the other fields these
columns give the inner, median, and maximum radius that the field
covers.  The sigmas of the Gaussians that best fit the bMS and rMS
color distributions, with their uncertainties, are in Columns 5--8.
Columns 9 and 10 give the \nbr\ ratio and its error.  Column 11 gives
the difference (in straightened color) between the peaks of the
Gaussians that best fit the bMS and rMS.  The last column gives the
color baseline of each data set.  By $\Delta {\rm color}$ we mean a
color difference or width, in the straightened CMD [either ($B$,
$B-R_C$) or ($m_{\rm F606W}$, $m_{\rm F606W}-m_{\rm F814W}$),
whichever applies].

Our results are qualitatively consistent with those of Sollima et al.\
(\cite{sollima07}), within the common region of radial coverage.  We
confirm the flat radial distribution of \nbr\ outside $\sim$8--10
arcmin, and a clear increase of \nbr\ toward the cluster center.  For
the first time, and as a complement to the Sollima et al.\
(\cite{sollima07}) investigation, our ACS/WFC 3$\times$3 mosaic data set
has enabled us to study the distribution of \om MS stars in the
innermost region of the cluster.  Inside of $\sim$1.5 \rc\ (i.e., inward
of $\sim$2$\arcmin$), the \nbr\ ratio is almost flat and close to unity,
with a slight overabundance of bMS stars.  At larger distances from the
cluster center, the \nbr\ ratio starts decreasing.  Between
$\sim$3$\arcmin$ and $\sim$8$\arcmin$ (the latter corresponding to
$\sim$2 half-mass radii) the ratio rapidly decreases to $\sim$0.4, and
remains constant in the cluster envelope.  Better azimuthal and radial
coverage of the region where the maximum gradient is observed would be
of great value.  In the radial interval between 1 and 2 half-mass radii,
we can use only the corners of the \iacs, and the FORS1 photometry,
which inside of $10\arcmin$ is seriously affected by crowding and
saturated stars.  In any case, the star counts and even visual
inspection of the histograms in Fig.~\ref{fig_dualGauss_fits} leave no
doubt about the overall gradient.

Note that in the two innermost bins there are more bMS than rMS stars,
even though the heights of the two peaks would suggest the opposite.
The apparent contradiction disappears, however, when we note the much
greater width of the bMS Gaussian, which more than makes up for the
difference in heights.  This seems to be consistent with a greater
spread in chemical composition for metal-intermediate than for
metal-poor stars, as first seen by Norris et al.\ (\cite{norris97}).
Our approach, using a dual-Gaussian fit, has been optimized to estimate
the value of the number ratio of bMS to rMS stars, avoiding as much as
possible any contamination by blends, binaries, and MS-a stars.

We must also address the fact that the \nbr\ values found by Sollima
et al.\ (\cite{sollima07}) are consistently lower than our values.
The difference is largely due to their use, on the red side, of a wide
color range (see their Fig.\ 5) that includes nearly all of the
contamination by blends, binaries, and MS-a stars that our method has
so studiously avoided.  This makes their numbers of rMS stars much too
high---easily enough to account for their finding a value of
$\sim$0.16 in the cluster envelope, rather than our $\sim$0.4, which
is certainly much closer to the truth.  Note also that we have
concentrated exclusively on the {\it ratio} of numbers of bMS and rMS
stars, making no attempt to derive absolute numbers for each
component.  We felt that absolute numbers would be subject to
different incompleteness corrections in our different data sets,
whereas the incompleteness in each data set should be the same for
each component and should therefore not affect their ratio.

Finally, the robustness of our method is shown by the close agreement
of our --- proper-motion selected --- outer ACS field (magenta open
circle in Fig.~\ref{fig_gradient}), which has almost no crowding
problems, with the outer ground-based FORS1 fields (last two red
squares in Fig.~\ref{fig_gradient}), which are certainly affected
somewhat by crowding.

%################################################################

\subsection{Artificial star tests}
\label{subsec:at}

\begin{table*}[th!]
\centering
\caption{Results of the two artificial-star tests.  For each of the
    two fields (first column), we give in Cols.\ 2--4 the values of
    \nbr\ for the AS that were inserted, and the color dispersions
    that were given to the AS that were put on the bMS and rMS,
    respectively. In Cols.\ 5--8 are, respectively, the \nbr\ of the
    AS that were recovered, with error, followed by the sigmas of the
    two Gaussians that were fitted to them.  See text for details.}
\label{tab:at}
\begin{tabular}{cccccccc}
\hline
\hline
& & & & & & & \\
Field & $({N_{\rm bMS}}/{N_{\rm rMS}})_{\rm ins.}$ &
        $({\sigma_{\rm bMS}})_{\rm ins.}$ & $({\sigma_{\rm rMS}})_{\rm ins.}$ &
        $({N_{\rm bMS}}/{N_{\rm rMS}})_{\rm rec.}$ & 
        $\sigma(N_{\rm bMS}/N_{\rm rMS})_{\rm rec.}$ &
$({\sigma_{\rm bMS}})_{\rm rec.}$ & $({\sigma_{\rm rMS}})_{\rm rec.}$ \\
& & & & & & & \\
\hline
& & & & & & & \\
\multicolumn{7}{c}{TEST1}\\
& & & & & & & \\
central & 1.000 & 0.000 & 0.000 & 0.946 & 0.013 & 0.013 & 0.013\\
corner  & 1.000 & 0.000 & 0.000 & 0.992 & 0.013 & 0.009 & 0.009\\
& & & & & & \\
\multicolumn{7}{c}{TEST2}\\
& & & & & & \\
central & 1.000 & 0.021 & 0.016 & 1.008 & 0.079 & 0.026 & 0.021\\
corner  & 1.000 & 0.016 & 0.013 & 0.996 & 0.027 & 0.019 & 0.016\\
& & & & & & \\
\hline
\end{tabular}
\end{table*}

\begin{figure*}[ht!]
\centering
\includegraphics[width=9cm]{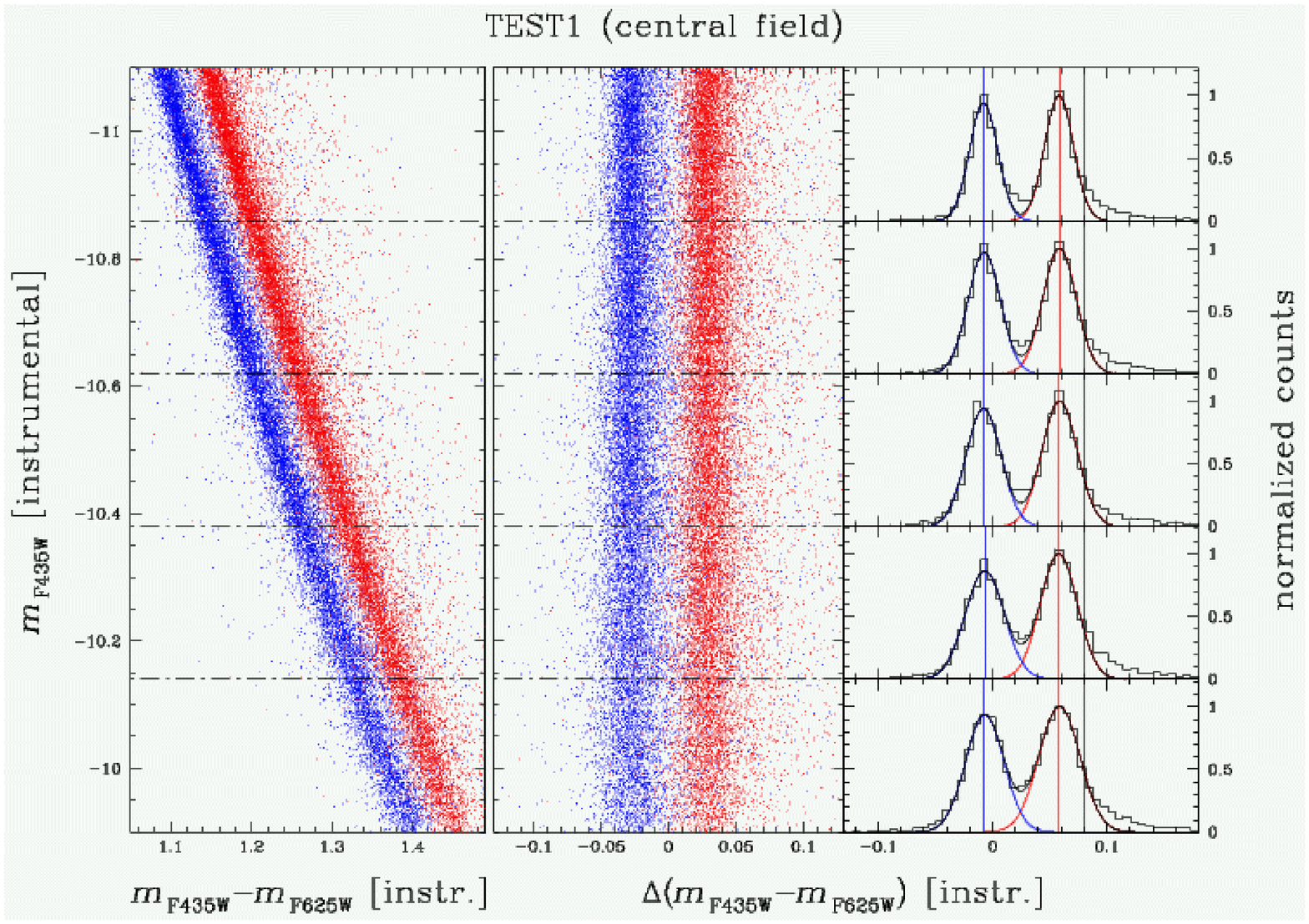}
\includegraphics[width=9cm]{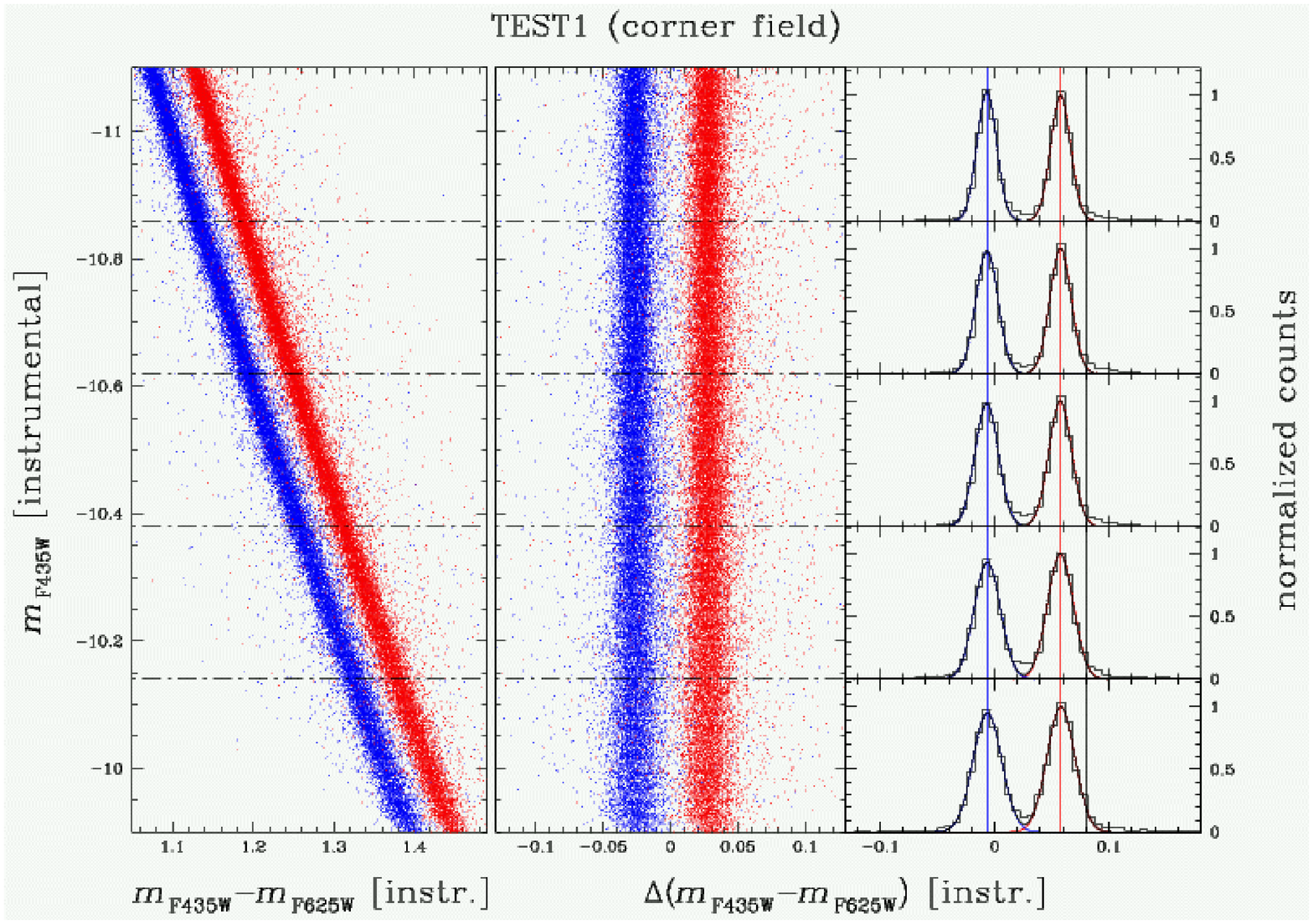}
\caption{TEST1 artificial star analysis for the central \iacs\
    field (left panels), and for the corner field (right panels).  For
    each panel, we show the CMD with the recovered stars (in blue for
    the bMS stars and in red for the rMS stars), for five magnitude
    intervals. The straightened MSs are plotted in the middle, while
    on the right we show the color histograms, with the dual-Gaussian
    fits. The vertical lines in blue, red, and black mark, respectively,
    the centers of the two Gaussians and the red cut-off.  See text for
    details.}
\label{fig:at1}
\end{figure*}

\begin{figure*}[ht!]
\centering
\includegraphics[width=9cm]{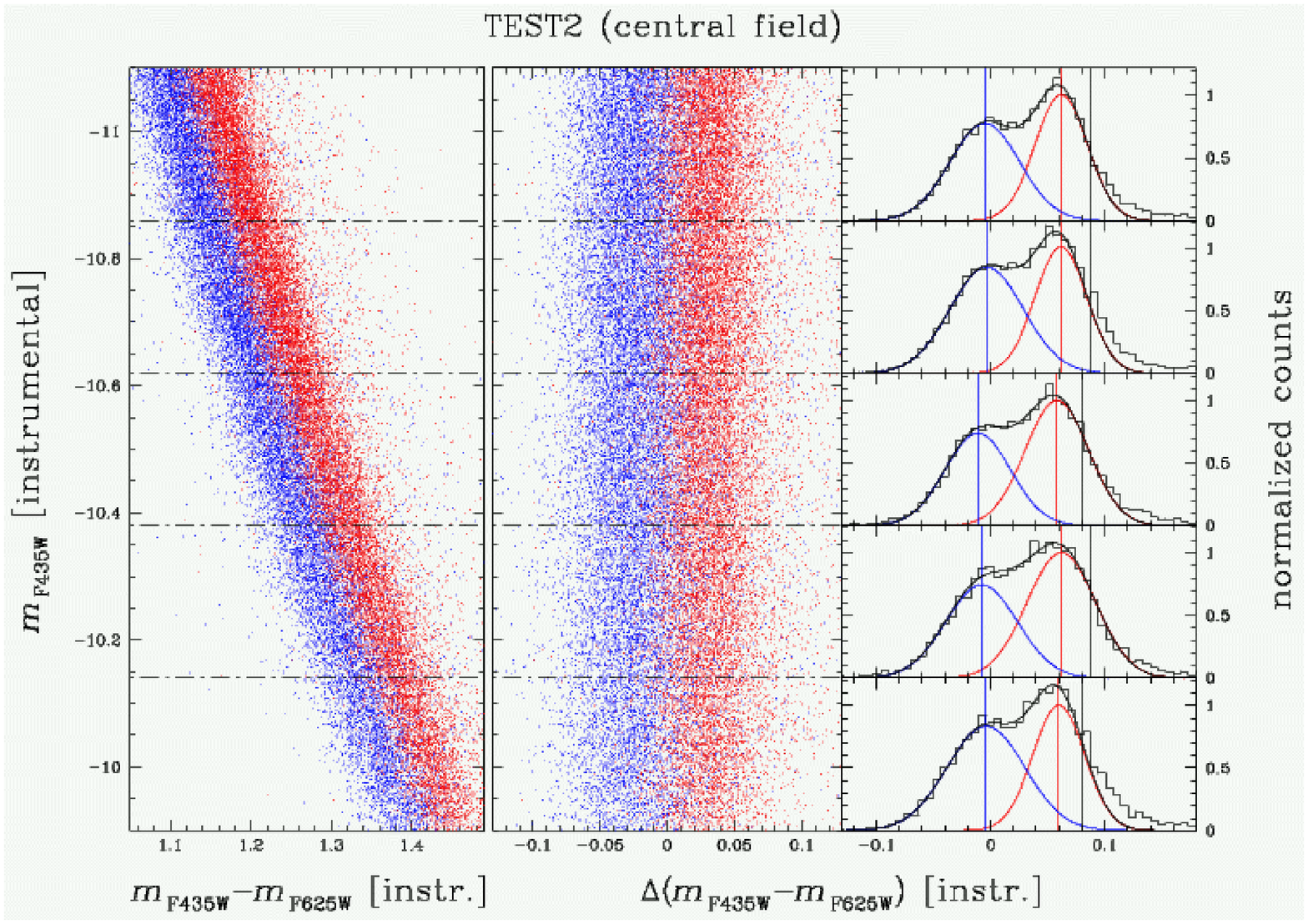}
\includegraphics[width=9cm]{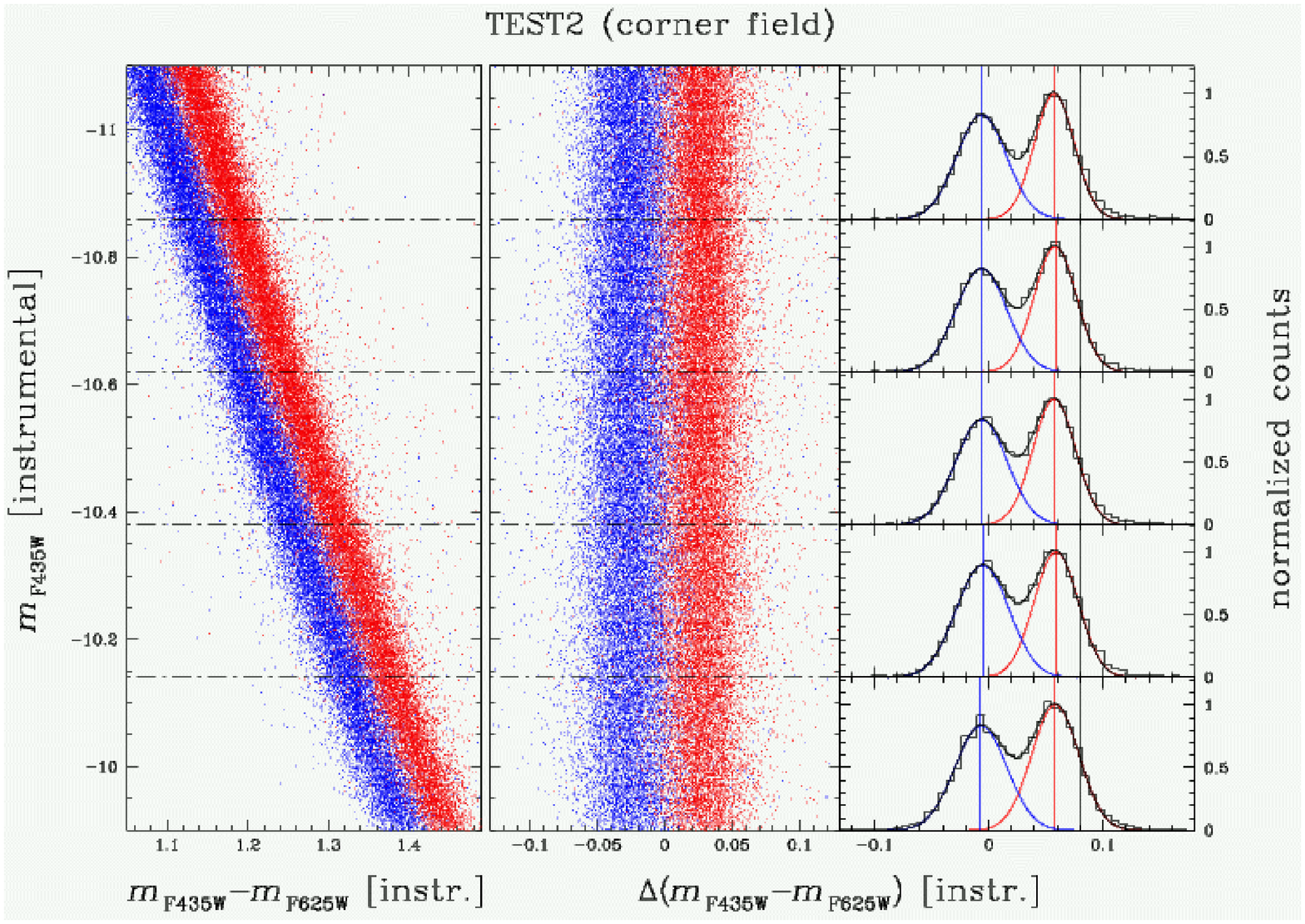}
\caption{Same as Fig.~\ref{fig:at1}, but now for TEST2. See text
for details.}
\label{fig:at2}
\end{figure*}

Even with the technique that we have used to exclude the effects
of photometric blends and binaries, which lie above and to the red of
any MS, there is a concern that some bMS stars would be shifted into the
rMS region (and some rMS stars lost on the red side of the MS), and that
these shifts would distort the \nbr\ ratio.  As a check against this
possibility we have made two tests using artificial stars (AS).  In each
test we introduced the same AS into both the F435W and the F625W images,
as follows.

For each test, we first created $45\,000$ artificial stars, with
random F435W instrumental magnitudes between $-$11.1 and $-$9.9
(corresponding to $20.9<B<21.1$), and random positions.  We then took
each of these $45\,000$ AS, assigned it a color that placed it on the
bMS, and inserted it in the F625W images, at the same position but
with the F625W magnitude that corresponds to this color.  We then
repeated this procedure for $45\,000$ new AS, but this time we gave
them colors that put them on the rMS.  (What we mean by ``on'' [bMS or
rMS] differs between the two tests; see below for an explanation of
the difference.)  Each artificial star in turn was added, measured,
and then removed, so as not to interfere with the other AS that were
to be added after it; this procedure is that of Anderson et al.\
(2008), where it is explained in detail.

In order to test the effects of crowding, each of the two
tests used two fields from the central 3$\times$3 mosaic:\ the central
field where crowding effects are maximal, and one of the corner
fields, about 5 arcmin ($3.6r_{\rm c}$) southeast of the center (see
Fig.\ \ref{fig_fow} for a map of the 3$\times$3 mosaic of fields).

The first of the tests (TEST1) was aimed at checking the
photometric errors in the colors.  To do this, we chose the color of
each AS so as to put it exactly on the ridge line of the bMS or rMS; the
color spread of the recovered AS would then serve as a lower-limit
estimate of our photometric error.

The aim of TEST2 is to verify our ability to insert AS with
\nbr=1 and then recover that value, when the two MSs have intrinsic
dispersions in color.  To do this, we first derived the intrinsic
spreads of the two sequences by taking from the fifth and seventh
columns of Table \ref{tab:MS} the simple unweighted mean of the entries
in lines 1 and 2 for the central field, and in lines 4 and 5 for the
corner one.  (The more fastidious procedure, weighting the entry in each
of the two lines according to the number of stars contributed by that
annulus, would have been quite laborious and would have made no
significant change in the results.)  These are the observed total color
spreads (intrinsic spread plus measuring error) of the bMS and rMS,
respectively, in the two fields that we are using here.  From these
total spreads we quadratically subtracted the corresponding
measuring-error spreads that we had found in TEST1, so as to get
estimates of the intrinsic color spreads of the two sequences.  We
created AS in the same manner as in TEST1, but this time instead of
placing the AS on the center lines of bMS and rMS, we adjusted the F625W
magnitudes so as to give the AS a Gaussian spread in color around each
sequence, using the intrinsic sigmas that we had just found.  After the
measuring process, these AS should duplicate the observed total spreads,
and can be used to estimate the amount of contamination between the two
main sequences.  To repeat, each test was performed both on both the
central and the corner field.

The results of these AS tests are summarized in brief numerical
form in Table~\ref{tab:at} and in graphical form in Figures
\ref{fig:at1} and \ref{fig:at2}.  In each figure the left and right
halves refer to the central and corner fields, respectively, while each
half figure is divided into three panels that show, from left to right,
the CMD, the straightened CMD, and the decomposition of the number
densities of the latter into best-fitting Gaussians.  Each panel showing
the Gaussian fits is subdivided into five magnitude intervals, (very
similarly to what is done in Fig.~8).  Cols.\ 2--4 of Table \ref{tab:at}
give, for each field and AS test, the \nbr\ ratio of the inserted AS and
the dispersions of the MSs. The recovered values (weighted mean of the
five magnitude bins and its error, as explained in detail for real stars
in Sect.\ 3.3) are shown in Cols.\ 5--8.

From the results of TEST1 we conclude that in each field the color
spread introduced by measuring error is the same for bMS stars as for
the rMS, and that it is about 40\% higher in the central field than in
the less-crowded corner field.  TEST1 has served two purposes: (1) It
gave us a clear, effective measure of the effect of crowding on the
color spread.  (2) It evaluated the color spreads due to measuring
error alone, which we used in setting up TEST2.  (Its results for
\nbr\ are given, pro forma, but they have no real significance, since
the color spreads used in TEST1 are so narrow that our color bin-width
does not sample them adequately.)  It is TEST2 which directly tests
our previous conclusions about the size of \nbr.  We conclude from it
that the AS tests recover our input values of \nbr, within the
uncertainties of the measurement.

In this section we have demonstrated, on two extreme fields of the ACS
inner mosaic, that our dual-Gaussian fitting method is fully effective
in overcoming the effects of crowding on the distribution of colors,
and that it reliably estimates the relative star numbers in the two
sequences.  (Note that we use this same method for all of our other
data sets too).  As we noted at the end of Sect. 3.3, the excellent
agreement between the results from our completely uncrowded outer ACS
field and those from our outer FORS1 fields establishes the validity
of the latter, without recourse to any additional AS texts for them.

%###################################################################

\section{Radial gradients in the RGB subpopulations}
\label{sec:rgb}

It has been known since the end of the 60s that the RGB of
\om is broader than would be expected from photometric errors
(Woolley \& Dickens \cite{woolley67}), but it was only in
1999 that Lee et al.\ (\cite{lee99}) clearly detected at least two
distinct RGBs.  Later on, Pancino et al.\ (\cite{pancino00})
demonstrated that there is a correlation between the photometric peaks
across the RGB and three peaks in the metallicity distribution.  On
this basis, they defined the three RGB groups:\ RGB-MP, RGB-MInt, and
RGB-a, characterized by an increasing metallicity.  In this section we
will present a detailed study of the radial distributions of these
components.

%###################################################################

\subsection{Defining the RGB-MP, RGB-MInt, and RGB-a subsamples}

Unfortunately the WFPC2, FORS1, and outer ACS/WFC data sets we used to
analyze the main-sequence population in the previous section are
saturated even at the MS turn-off level, and are therefore unusable
for the study of the RGB radial distributions.

Our \wfi\ photometric and proper-motion catalog (Bellini et al.\
\cite{bellini09}), however, is an excellent data base for this study,
particularly in view of the fact that we can safely remove field
objects in the foreground and background, thanks to our accurate
proper motions.  This proper-motion cleaning is of fundamental
importance in the outer envelope of the cluster, where there can be
more field stars than cluster giants.  In the central regions of the
cluster, the \wfi\ data are less accurate due to the poorer photometry
caused by the crowded conditions, so there we take advantage of our
high-resolution inner \iacs, which included short exposures to measure
the bright stars.  Below we describe how we extracted the \om RGB
subsamples from these two data sets.

Because of the complex distribution of the stars along the RGB we were
forced to use bounding boxes to select the different RGBs.  This
selection procedure is less accurate than what we were able to do for
the bMS and the rMS; nevertheless it is still accurate enough to study
the general trend of the radial distribution of the relative numbers
of RGB-MP, RGB-Mint, and RGB-a stars.  The Poisson error from the
smaller number of RGB stars makes the more precise procedure less
critical.

For the ACS data, we defined bounding boxes for the RGB subpopulations
of \om\ in the CMD obtained from the data set of the \iacs, for which
the large-number statistics make the separation among the different
RGBs easier to see.  We extracted three RGB subpopulations, in a way
very similar to that used by Ferraro et al.\ (\cite{ferraro02}).
[Note that other authors (e.g.\ Rey et al.\ \cite{rey04}; Sollima et
al.\ \cite{sollima05a}, Johnson et al.\ \cite{johnson09}) have defined
four or even five RGB subpopulations].  The left panel of
Fig.~\ref{fig:cmds_rgb} shows the three RGB bounding-box regions drawn
in the CMD from the \iacs, to identify the three subgroups RGB-MP,
RGB-MInt, and RGB-a.  Our RGB selections are limited to magnitudes
brighter than $B=17.9$, and contain 5184 RGB-MP stars, 4379 RGB-MInt
stars, and 383 RGB-a stars.

In extracting the RGB subpopulations from our \wfi\ data set we chose
to define the subpopulations in the $B$, $B-V$ CMD. Even though we
cannot adopt exactly the same selection boxes in the $B$, $B-R_C$ CMD
as for the \iacs.  This choice might appear awkward, not only because
the color baseline $B-V$ is shorter than the $B-R_C$ baseline, but
also because the WFI $R_C$ filter is very similar to the ACS/WFC
$F625W$ filter.  There are other good reasons for adopting the $B-V$
color baseline, however.  The most important one is that the WFI
photometry obtained with the $V$ filter has ten times as much
integration time, and more dithered images than those available for
the $R_C$ filter.  Therefore our $V$ photometry is considerably more
precise, and more accurate, than our $R_C$ magnitudes.  Moreover, our
empirical sky-concentration correction (very important for such
studies) is better defined in $V$ than in $R_C$ (see Bellini et al.\
\cite{bellini09}).

In this \wfi\ $B$ vs.\ $B-V$ CMD, we tried to define the bounding
boxes in a way that was as consistent as possible with what we did for
the data set from \iacs.  We cross-identified the stars that are in
common between the sample that we had selected from the RGB CMD of the
\iacs, on the one hand, and the \wfi\ $B-V$ data set on the other
hand, and we carefully drew by hand, in the $(B,B-V)$ CMD, bounding
boxes that would include the same stars as in the sample from the
\iacs.

In addition, we selected from the \wfi\ data set the stars that were
measured best (both photometrically and astrometrically), and were
most likely to be members of \oma.  To make the selection we used the
error quantities in columns 7, 9, 13, and 15 of Table 6 of Bellini et
al.\ (\cite{bellini09}).  These are the errors of the two components
of proper motion and of the $B$ and $V$ magnitudes.  Our selection
consisted of choosing, at the bright end of the RGB, stars whose
proper-motion error has a magnitude less than 1.8 \masyr, and whose
photometric error is less than 0.02 mag in each band,; we also
required that the proper motion of a star differs from the mean motion
of cluster stars by no more than 2.1 \masyr.  At the faint end of the
RGB we allowed these three tolerances to rise to: 2.1 \masyr, 0.03 mag
and 3.8 \masyr, respectively.  This high-quality data set comprised
4993 RGB-MP stars, 3057 RGB-MInt, and 292 RGB-a stars.

The right-hand panel of Fig.~\ref{fig:cmds_rgb} shows the \wfi\ RGB
subpopulations that were selected in this way.  We note that whereas the
RGB-a sample is well separated from the other two RGB components, the
RGB-MP and RGB-MInt components are separated only by an arbitrary
dividing line, so that small differences in defining the bounding boxes
might result in some cross-contamination in those two samples.

%##############################################################

\subsection{Relative radial distributions of RGB stars}
\label{sec:rgbrad}
 
\begin{figure}[t!]
\centering
\includegraphics[width=9cm]{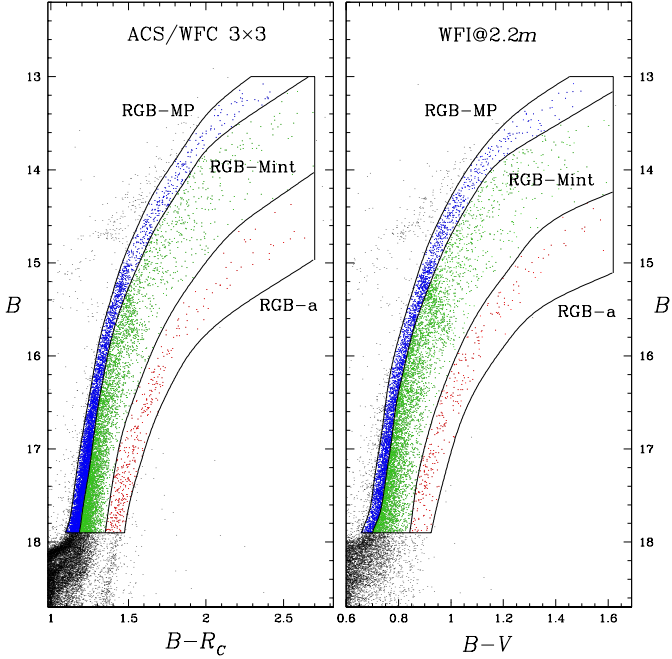}
\caption{CMDs of the \om RGB from \iacs\ ($B$ vs.\  $B-R_C$,
          \textit{left panel}) and from \wfi\ data ($B$ vs.\ $B-V$,
          \textit{right panel}).  The RGB subpopulations selected are
          also plotted with different colors.  See text for details.}
\label{fig:cmds_rgb}
\end{figure}

\begin{figure}[t!]
\centering
\includegraphics[width=9cm]{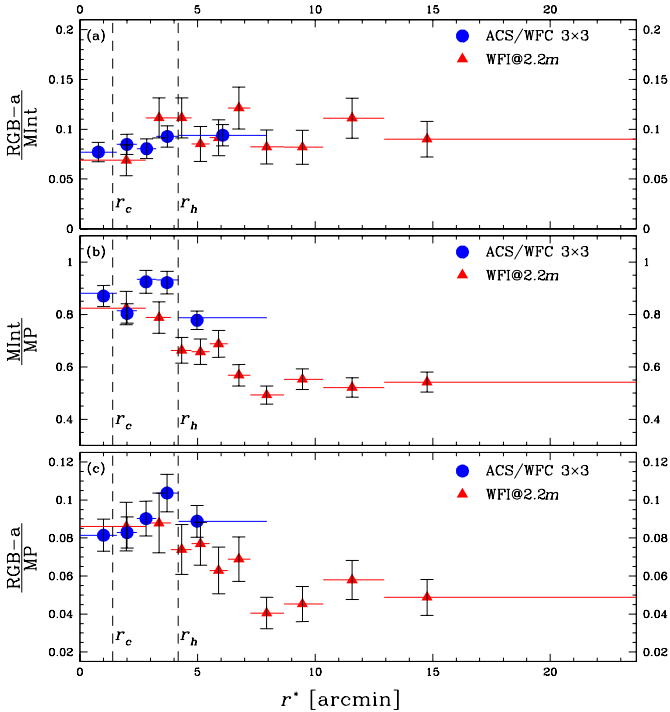}
\caption{\textit{(a)}:\ Radial distribution of the ratio
                      RGB-a/RGB-MInt for the \wfi\ data set (red triangles)
                              and for the ACS/WFC 3$\times$3 data set (blue
                              circles).  Vertical dashed
                              lines mark the core radius and the half-mass
                              radius, respectively.
          \textit{(b)}:\ Radial distribution of the ratio
                              RGB-MInt/RGB-MP.
          \textit{(c)}:\ Radial distribution of the ratio RGB-a/RGB-MP.
          See text for details.} 
\label{fig:gradients_rgb}
\end{figure}

We divided our \wfi\ data set into ten radial bins, each containing
approximately the same number of RGB-MInt stars, and the ACS/WFC
3$\times$3 data set into five radial bins, again with the same
equal-number criterion.  For each of these bins we counted the number
of RGB stars in each subpopulation.

In Fig.~\ref{fig:gradients_rgb} we show the derived radial gradients.
As it has not been possible to perform the same error analysis as was
done for the MS stars (because of the much smaller number of stars),
the error bars in Fig.~\ref{fig:gradients_rgb} represent only Poisson
errors, and should be considered a lower limit to the real errors.  In
panel (a) we show the radial distribution of the ratio RGB-a/RGB-MInt.
Blue full circles refer to the ACS/WFC 3$\times$3 data set, and red
triangles to the \wfi\ data.  Vertical dashed lines mark the core
radius \rc\ and the half-mass radius \rh.  We found that, within the
errors, the RGB-a and the RGB-MInt stars share the same radial
distribution, since their ratio is constant over the entire radial
range covered by our two data sets.  In panel (b), we plot the ratio
RGB-MInt/RGB-MP for the two data sets.  The RGB-MInt stars are more
centrally concentrated than the RGB-MP stars, with a flatter trend
within $\sim$1 \rh, a rapid decline out to
$\sim$8$\arcmin-$10$\arcmin$, and again a flat relative distribution
outside.  There is a hint, also, that the RGB-MInt/RGB-MP ratio could
be nearly constant within the half-mass radius.  We find that the
general radial trend of the RGB-MInt/RGB-MP star-count ratio is
consistent with that of \nbr.  This result provides additional
evidence (in agreement with the metallicity measurements by Piotto et
al.\ \cite{piotto05}) that the bMS and the RGB-MInt population must be
part of the same group of stars, with the same metal content and the
same radial distribution within the cluster.  Panel (c) shows that the
ratio RGB-a/RGB-MP resembles, within the errors, the RGB-MInt/RGB-MP
trend.  We were unable to examine this trend for the MS part of the
RGB-a population, since the MS-a sequence cannot be followed below B
$\sim$20.

Our analysis confirms the results by Norris et al.\ (\cite{norris97}),
Hilker \& Richtler (\cite{hilker00}), Pancino et al.\
(\cite{pancino00}), and Rey et al.\ (\cite{rey04}), and Johnson et
al.\ (\cite{johnson09}), who found that the most metal-poor RGB stars
are less concentrated than the RGB-MInt ones.  Moreover, we can also
confirm that the RGB-a and the RGB-MInt share the same radial
distribution within $\omega$\ Cen, as found by Norris et al.\
(\cite{norris97}), Pancino et al.\ (\cite{pancino00}),
 and Pancino et al. (\cite{pancino03}) for RGB-a only.

It is important to note that because we were able to use proper
motions to construct a pure cluster sample, our results are not
affected by field-star contamination, which would tend to enhance the
RGB-a star counts in the cluster outskirts with respect to the more
populous RGB-MP sample (which also covers a smaller region in the
CMD).  Field-star contamination is likely the reason that Hilker \&
Richtler (\cite{hilker00}) and Castellani et al.\
(\cite{castellani07}) found the RGB-a/RGB-MP ratio to {\it increase}
with distance from the cluster center --- the opposite trend from what
is seen here.  
Moreover, it is interesting to note that the different
RGB-Mint subgroups (as highlighted, e.g., by Sollima et al.\
\cite{sollima05a}) might well have a different radial behavior, but
necessarily---since we cannot distinguish them in the CMD---we have to
treat them together and study only their average gradient.

%#########################################################################

\section{Discussion}

In this paper we have analyzed the radial distribution of the
different MS and RGB components in the globular cluster $\omega$\
Centauri.  We used high-resolution ACS/WFC images to study the inner
regions of the cluster, and ACS/WFC, WFPC2 and FORS1@VLT images, as
well as WFI@2.2$m$ images, for the cluster envelope.  We found that
there are slightly more bMS stars than rMS stars in the inner 2 core
radii.  At larger distances from the cluster center, out to $\sim$8
arcmin, the relative number of \nbr\ stars drops sharply, and then
remains constant at \nbr$\sim$0.4, out to half the tidal radius of the
cluster.

Our most precise photometry comes from the outer ACS field at
17\arcmin\ (12 \rc), where we find that the color dispersion
($\sigma$) of the bMS is about 50\% larger than that of the rMS.  The
other observations are consistent with this, though they are unable to
measure $\sigma$ so precisely, on account of crowding (in the inner
ACS field) and other errors (in the ground-based fields).

The RGB-MInt population (associated with the bMS by Piotto et al.\
\cite{piotto05}) and the RGB-MP sample (which includes the progeny of
the rMS) follow a trend similar to that of \nbr.  The most metal-rich
component of the RGB, RGB-a, also follows the same distribution as the
RGB-MInt component.

On the hypothesis that the bMS, the presumably helium-rich population,
is a second generation of stars formed by the low-velocity material
ejected by a primordial population (which we assume to be the more
metal-poor rMS population), the bMS must have formed from matter that
collected in the cluster center via some kind of cooling flow.  This
is in qualitative agreement with the recent models by Bekki \& Norris
(\cite{bekki06}) and D'Ercole et al.\ (\cite{d'ercole08}).  The very
long relaxation time (half-mass relaxation time longer than 10 Gyr,
according to the Harris \cite{harris96} compilation) has preserved
some information about the original kinematic and spatial distribution
of the material from which the younger component took form.
Interestingly enough, the third, most-metal-rich population is also
more concentrated than the most metal-poor component, and has a radial
distribution that is rather similar to that of the
intermediate-metallicity sample.  It is also noteworthy that the bMS
component has a broader color distribution than the rMS one.  This
fact may reflect, at least in part, the large dispersion in iron
abundance of the intermediate-metallicity component (e.g. Norris \& Da
Costa \cite{norris95}).  Alternatively, this bMS spread could be an
indication of the dispersion of other chemical elements, including He.
Only a detailed analysis of the metal content of the two MSs can solve
this issue, but for this we might need to wait for the next generation
of 30+ meter telescopes, on account of the faintness of these stars.

%#########################################################################

\begin{acknowledgements}

A.B.\ acknowledges support by the CA.RI.PA.RO.\ foundation, and by
STScI under the 2008 graduate research assistantship program.  I.R.K.\
and J.A.\ acknowledge support by STScI under grants GO-9442, GO-9444,
and GO-10101.  G.P.\ and A.P.M.\ acknowledge partial support by MIUR
under the program PRIN2007 (prot.\ 20075TP5K9) and by ASI under the
program ASI-INAF I/016/07/0.

\end{acknowledgements}

%#########################################################################

{}


\begin{thebibliography}{}

\bibitem[1997]{anderson97} Anderson, J., Ph.D.\ thesis, Univ.\ of
    California, Berkeley, 1997

\bibitem[2000]{anderson00} Anderson, J., \& King, I.~R. 2000, \pasp,
112, 1360

\bibitem[2002]{anderson02} Anderson, J.\ 2002, in Omega Centauri, A
Unique Window into Astrophysics, ed.\ F.\ van Leeuwen, J.\ D.\ Hughes,
\& G.\ Piotto, ASP Conf.\ Ser., 265 (San Francisco:\ ASP), p.\ 87

\bibitem[2006]{anderson06a} Anderson, J., Bedin, L.~R., Piotto, G.,
Yadav, R.~S., \& Bellini, A.\ 2006, \aap, 454, 1029

\bibitem[2006]{anderson06b} Anderson, J., \& King, I.~R.\ 2006, ACS/ISR
    2006-01 (Baltimore: STSci)

\bibitem[2008]{anderson08} Anderson, J., et al.\ 2008, \aj, 135, 2055

\bibitem[2009]{anderson09a} Anderson, J., Piotto, G., King, I.~R.,
Bedin, L.~R., \& Guhathakurta, P. 2009, \apjl, 697, L58

\bibitem[2009]{anderson09b} Anderson, J., \& van der Marel, R.~P.\
    2009, arXiv:0905.0627

%---------------------B
\bibitem[2004]{bedin04} Bedin, L.~R., Piotto, G., Anderson, J.,
Cassisi, S., King, I.~R., Momany, Y., \& Carraro, G.  2004, \apjl,
605, L125

\bibitem[2005]{bedin05} Bedin, L.~R., Cassisi, S., Castelli, F.,
Piotto, G., Anderson, J., Salaris, M., Momany, Y., \& Pietrinferni, A.
2005, \mnras, 357, 1038

\bibitem[2008]{bellazzini08} Bellazzini, M., et al.\ 2008, \aj, 136,
1147

\bibitem[2009]{bellini09} Bellini, A., et al.\ 2009, \aap, 493, 959

\bibitem[2006]{bekki06} Bekki, K., \& Norris, J.~E.\ 2006, \apjl, 637,
L109

\bibitem[2009]{bekki09} Bekki, K., \& Mackey, A.~D.\ 2009, \mnras,
394, 124

\bibitem[1976]{bessell76} Bessell, M.~S., \& Norris, J.\ 1976, \apj,
208, 369

\bibitem[1978]{butler78} Butler, D., Dickens, R.~J., \& Epps, E.\
1978, \apj, 225, 148

%---------------------C
\bibitem[2005]{calamida05} Calamida, A., et al.\ 2005, \apjl, 634, 
L69

\bibitem[1973]{cannon73} Cannon, R.~D., \& Stobie, R.~S.\ 1973, \mnras,
    162, 207

\bibitem[2006]{carretta06} Carretta, E., Bragaglia, A., Gratton,
R.~G., Leone, F., Recio-Blanco, A., \& Lucatello, S.\ 2006, \aap, 450,
523

\bibitem[2008]{carretta08} Carretta, E., Bragaglia, A., Gratton,
    R.~G., Lucatello, S.\ 2008, arXiv:0811.3591v1

\bibitem[2007]{castellani07} Castellani, V., et al.\ 2007, \apj, 663,
    1021

%---------------------D
\bibitem[2004]{dantona04} D'Antona, F., \& Caloi, V.\ 2004, \apj, 611,
871

\bibitem[2005]{dantona05} D'Antona, F., Bellazzini, M., Caloi, V.,
Fusi Pecci, F., Galleti, S., \& Rood, R.~T.\ 2005, \apj, 631, 868

\bibitem[2008]{d'ercole08} D'Ercole, A., Vesperini, E., D'Antona, F.,
McMillan, S.~L.~W., \& Recchi, S.\ 2008, \mnras, 391, 825


\bibitem[2008]{decressin08} Decressin, T., Baumgardt, H., \& Kroupa,
P.\ 2008, \aap, 492, 101

\bibitem[1999]{dinescu99} Dinescu, D.~I., Girard, T.~M., \& van
Altena, W.~F.\ 1999, \aj, 117, 1792

%---------------------F

\bibitem[2007]{faria07} Faria, D., Johnson, R.~A., Ferguson, A.~M.~N.,
Irwin, M.~J., Ibata, R.~A., Johnston, K.~V., Lewis, G.~F., \& Tanvir,
N.~R.\ 2007, \aj, 133, 1275


\bibitem[2002]{ferraro02} Ferraro, F.~R., Bellazzini, M., \& Pancino,
E.\ 2002, \apjl, 573, L95

\bibitem[2004]{ferraro04} Ferraro, F.~R., Sollima, A., Pancino, E.,
Bellazzini, M., Straniero, O., Origlia, L., \& Cool, A.~M.\ 2004,
\apjl, 603, L81

\bibitem[1975]{freeman75} Freeman, K.~C., \& Rodgers, A.~W.\ 1975,
    \apjl, 201, L71

\bibitem[1993]{freeman93} Freeman, K.~C.\ 1993, in The Globular
    Cluster--Galaxy Connection, ed.\ G.~H.\ Smith \& J.~P.\ Brodie
    ASP Conf.\ Ser., 48 (San Francisco:\ ASP), p,\ 608

\bibitem[2005]{frey05} Freyhammer, L.~M., et al.\ 2005, \apj, 623, 860

%---------------------G

\bibitem[1983]{geyer83} Geyer, E.~H., Nelles, B., \& Hopp, U.\ 
1983, \aap, 125, 359

\bibitem[2004]{gilliland04} Gilliland, R.\ 2004, ACS/ISR 2004-01
    (Baltimore: STScI)

%---------------------H
\bibitem[1996]{harris96} Harris, W.~E.\ 1996, \aj, 112, 1487, as
    updated in February, 2003.

\bibitem[2000]{hilker00} Hilker, M., \& Richtler, T.\ 2000, \aap, 362,
    895

\bibitem[2004]{hilker04} Hilker, M., Kayser, A., Richtler, T., \&
Willemsen, P.\ 2004, \aap, 422, L9


\bibitem[1995]{holtzman95} Holtzman, J.~A., Burrows, C.~J., Casertano,
S., Hester, J.~J., Trauger, J.~T., Watson, A.~M., \& Worthey, G.\
1995, \pasp, 107, 1065

%---------------------I
\bibitem[2004]{ideta04} Ideta, M., \& Makino, J.\ 2004, \apjl, 616, L107

%---------------------J
\bibitem[2009]{johnson09} Johnson, C.~I., Pilachowski, C.~A., Rich,
R.~M., \& Fulbright, C.~P.\ 2009, \apj, 698, 2048

%---------------------L
\bibitem[1999]{lee99} Lee, Y.-W., Joo, J.-M., Sohn, Y.-J., Rey, S.-C.,
Lee, H.-C., \& Walker, A.~R.\ 1999, \nat, 402, 55

%---------------------M
\bibitem[1991]{makino91} Makino, J., Akiyama, K., \& Sugimoto, D.\
1991, \apss, 185, 63

\bibitem[2001]{manfroid01} Manfroid, J., \& Selman, F.\ 2001, The
Messenger, 104, 16

\bibitem[2008]{marino08} Marino, A.~F., Villanova, S., Piotto, G.,
Milone, A.~P., Momany, Y., Bedin, L.~R., \& Medling, A.~M.\ 2008,
\aap, 490, 625

\bibitem[2009]{marino09} Marino, A.~F., Milone, A.~P., Piotto, G.,
Villanova, S., Bedin, L.~R., Bellini, A., \& Renzini, A.\ 2009, \aap
in press, arXiv:0905.4058

\bibitem[2008]{milone08} Milone, A.~P., et al.\ 2008, \apj, 673, 241

\bibitem[2009]{milone09} Milone, A.~P., Bedin, L.~R., Piotto, G., \&
Anderson, J. 2009, \aap, 497, 755

\bibitem[2009]{moretti09} Moretti, A., et al.\ 2009, \aap, 493, 539

%---------------------N
\bibitem[1975]{norris75} Norris, J., \& Bessell, M.~S.\ 1975, \apjl,
201, L75

\bibitem[1977]{norris77} Norris, J., \& Bessell, M.~S.\ 1977, \apjl,
211, L91

\bibitem[1995]{norris95} Norris, J.~E., \& Da Costa, G.~S.\ 1995,
\apjl, 441, L81

\bibitem[1996]{norris96} Norris, J.~E., Freeman, K.~C., \& Mighell,
K.~J.\ 1996, \apj, 462, 241

\bibitem[1997]{norris97} Norris, J.~E., Freeman, K.~C., Mayor, M., \&
Seitzer, P.\ 1997, \apjl, 487, L187

\bibitem[2004]{norris04} Norris, J.~E.\ 2004, \apjl, 612, L25

%---------------------P
\bibitem[2000]{pancino00} Pancino, E., Ferraro, F.~R., Bellazzini, M.,
Piotto, G., \& Zoccali, M.\ 2000, \apjl, 534, L83

\bibitem[2003]{pancino03} Pancino, E., Seleznev, A., Ferraro, F.~R.,
Bellazzini, M., \& Piotto, G.\ 2003, \mnras, 345, 683

\bibitem[2007]{pancino07} Pancino, E., Galfo, A., Ferraro, F.~R., \&
Bellazzini, M.\ 2007, \apjl, 661, L155

\bibitem[2005]{piotto05} Piotto, G., et al.\ 2005, \apj, 621, 777

\bibitem[2007]{piotto07} Piotto, G., et al.\ 2007, \apjl, 661, L53

\bibitem[2009]{piotto09} Piotto, G.\ 2009, in The Ages of Stars, ed.\
    E.~E.\ Mamajek, D.~R.\ Soderblom, \& R.~F.~G.\ Wyse, IAU Symposium
    No.\ 258 (Cambridge:\ Cambridge University Press), p.\ 233

\bibitem[2003]{platais03} Platais, I., Wyse, R.~F.~G., Hebb, L., Lee,
    Y.-W., \& Rey, S.-C.\ 2003, \apj, 591, L127

%---------------------R
\bibitem[1988]{renzini88} Renzini, A., \& Fusi Pecci, F.\ 1988, \araa,
26, 199

\bibitem[2008]{renzini08} Renzini, A.\ 2008, \mnras, 391, 354

\bibitem[2004]{rey04} Rey, S.-C., Lee, Y.-W., Ree, C.~H., Joo, J.-M.,
Sohn, Y.-J., \& Walker, A.~R.\ 2004, \aj, 127, 958

\bibitem[2004]{rich04} Rich, R.~M., Reitzel,
D.~B., Guhathakurta, P., Gebhardt, K., \& Ho, L.~C.\ 2004, \aj, 127, 2139 


%---------------------S
\bibitem[1995]{ata95} Sarajedini, A., \& Layden, A.~C.\ 1995, \aj, 109, 1086

\bibitem[2007]{sarajedini07} Sarajedini, A., et al.\ 2007, \aj, 133,
1658

\bibitem[1977]{searle77} Searle, L.\ 1977, in Evolution of Galaxies
and Stellar Populations, ed.\ B.~M.\ Tinsley, \& R.~B.\ Larson (New
Haven:\ Yale Univ.\ Obs.), p.\ 219

\bibitem[2007]{siegel07} Siegel, M.~H., et al.\ 2007, \apjl, 667, L57

\bibitem[2005]{sirianni05} Sirianni, M., et al.\ 2005, \pasp, 117,
1049

\bibitem[1987]{stetson87} Stetson, P.~B.\ 1987, \pasp, 99, 191

\bibitem[1994]{stetson94} Stetson, P.~B.\ 1994, \pasp, 106, 250

\bibitem[2000]{stetson00} Stetson, P.~B.\ 2000, \pasp, 112, 925

\bibitem[2005]{stetson05} Stetson, P.~B.\ 2005, \pasp, 117, 563

\bibitem[2005a]{sollima05a} Sollima, A., Ferraro, F.~R., Pancino, E.,
\& Bellazzini, M.\ 2005, \mnras, 357, 265 (2005a)

\bibitem[2005b]{sollima05b} Sollima, A., Pancino, E., Ferraro, F.~R.,
Bellazzini, M., Straniero, O., \& Pasquini, L.\ 2005, \apj, 634, 332
(2005b)

\bibitem[2007]{sollima07} Sollima, A., Ferraro, F.~R., Bellazzini, M.,
Origlia, L., Straniero, O., \& Pancino, E.\ 2007, \apj, 654, 915

\bibitem[2006]{stanford06} Stanford, L.~M., Da
Costa, G.~S., Norris, J.~E., \& Cannon, R.~D.\ 2006, \apjl, 653, L117

\bibitem[1996]{sunt96} Suntzeff, N.~B., \& Kraft, R.~P.\ 1996, \aj, 111, 
1913

%---------------------T

\bibitem[2004]{tsuchiya04} Tsuchiya, T.,
Korchagin, V.~I., \& Dinescu, D.~I.\ 2004, \mnras, 350, 1141 

%---------------------V
\bibitem[2006]{vdv06} van de Ven, G., van den Bosch, R.~C.~E.,
Verolme, E.~K., \& de Zeeuw, P.~T.\ 2006, \aap, 445, 513

\bibitem[2009]{vdm09} van der Marel, R.~P., \& Anderson, J.\ 2009,
arXiv:0905.0638

\bibitem[2007]{villanova07} Villanova, S., et al.\ 2007, \apj, 663,
296

%---------------------W
\bibitem[1967]{woolley67} Woolley, R.~V.~d.~R., \& Dickens, R.~J.\
  1967, Roy.\ Obs.\ Bull., No.\ 128

%---------------------Z
\bibitem[1988]{zinnecker88} Zinnecker, H., Keable, C.~J., Dunlop,
   J.~S., Cannon, R.~D., \& Griffiths, W.~K.\ 1988, in The
   Harlow-Shapley Symposium on Globular Cluster Systems in Galaxies, ed.\
   J.~E.\ Grindlay \& A.~G.~D.\ Philip, IAU Symposium No.\ 126
   (Cambridge:\ Cambridge University Press), p.\ 603

\end{thebibliography}
\end{document}